\documentclass[useAMS,usegraphicx,usenatbib,times,hyperref]{mn2e}
\usepackage[usenames]{color}
\definecolor{grey}{rgb}{0.35,0.35,0.35}
\definecolor{dblue}{rgb}{0.05,0.05,0.35}

\usepackage[ps2pdf,colorlinks=true,breaklinks=true,bookmarks=true,hypertexnames=false,bookmarksnumbered=true,backref=page,pagebackref=true,citecolor= dblue,linkcolor= grey,bookmarksopen=false,bookmarks=true,bookmarksnumbered=true]{hyperref}

\usepackage{graphicx}
\usepackage{subfigure}
\usepackage{amsmath,amssymb}
\usepackage{algorithm}
\usepackage{algorithmicx}
\usepackage{algpseudocode}

\def\progname{DisPerSE\,}
\def\paperone{Sousbie  (2010)}

\newcommand{\kpn}[2][k]{( #1 \!+\! #2 )}
\newcommand{\kmn}[2][k]{( #1 \!-\! #2 )}
\def\kp1{$\kpn{1}$}
\def\km1{$\kmn{1}$}

\def\Mpc {{\,h^{-1}\,{\rm Mpc}}}
\def\Mpccc {{\,h^{-3}\,{\rm Mpc}^{3}}}

\def\and  {\it {et al.} \rm}

\def\lcdm {\Lambda{\rm CDM}}
\def\nsig#1{{\rm #1-}\sigma}
\def\deg{^{\circ}}

\def\ppair#1#2{\left[#1,#2\right]}

\def\eg {{\it e.g. }}
\def\ie {{\it i.e. }}
\def\ni{\noindent}

\def\rhou{{\rho_\uparrow}}
\def\rhod{{\rho_\downarrow}}

\usepackage[usenames]{color}
\definecolor{Blue}{rgb}{0,0.08,0.65}
\definecolor{Red}{rgb}{0.65,0.08,0.05}
\definecolor{Green}{rgb}{0.15,0.65,0.25}

\begin{document}
\title[Persistent cosmic web II: Illustrations ]{The persistent cosmic web and its filamentary structure \\ II: 
Illustrations}
\author[T. Sousbie, C. Pichon, H. Kawahara]
{\ni T. Sousbie$^{1,2}$\thanks{{tsousbie@gmail.com} }, \ni C. Pichon$^{2,3}$, \ni H. Kawahara$^{4}$\\
$^{1}$Department of Physics, The University of Tokyo, Tokyo 113-0033, Japan\\
 $^{2}$Institut d'astrophysique
de Paris \& UPMC (UMR 7095), 98, bis boulevard Arago , 75 014, Paris,
France.\\
$^{3}$Oxford Astrophysics, department of Physics,
Denys Wilkinson Building,  Keble Road,
Oxford OX1 3RH, UK.\\
$^{4}$Department of Physics, Tokyo Metropolitan University, Hachioji, Tokyo 192-0397, Japan
}
\maketitle
% =====================================================================
\begin{abstract}

The recently introduced discrete persistent structure extractor (\progname\,  Soubie  2010, paper I) is implemented on realistic 3D cosmological simulations and observed  redshift catalogues; it is found that \progname\, traces  very well the observed filaments, walls, and voids seen both in simulations and observations. In either setting, filaments are shown to connect onto halos, outskirt  walls, which circumvent voids, as is topologically required  by Morse theory. Indeed this algorithm returns the optimal  critical set while operating directly on the particles. \progname, as illustrated here, assumes nothing about the geometry of the survey  or its homogeneity, and yields a natural (topologically motivated) self-consistent criterion for selecting the significance level of the identified structures. It is shown  that this extraction  is possible even for very sparsely sampled  point processes,  as a function of the persistence ratio (a measure of the significance of topological  connections between  critical points). Hence astrophysicists should be in a position to trace precisely the locus of filaments, walls and voids from such samples and  assess the  confidence of the post-processed sets as a function of this threshold, which can be expressed relative to the expected amplitude of  shot noise. In a cosmic framework, this criterion is shown to level with  friend of friend structure finder for the identifications of peaks, while it also identifies the connected filaments  and walls, and quantitatively recovers the  full set of topological invariants (number of holes, etc...) {\sl directly from the particles}, and at no extra cost as a function of the persistence threshold. This criterion is found to be sufficient even if one particle out of two is noise,  when the  persistence ratio is set to 3-sigma or more. The algorithm is also implemented on the SDSS catalogue and used to locat interesting configurations of the filamentary structure. In this context we carried the identification of an ``optically faint'' cluster at the intersection of filaments through the recent observation of its X-ray counterpart by SUZAKU. The corresponding filament catalogue is available online.
\end{abstract}
% =====================================================================
% =====================================================================

%=====================
\section{Introduction}
%=====================
\label{sec:intro}

Over the past decades, numerical simulations \citep[\eg][]{Efstathiou},  and large redshift surveys \citep[\eg][]{lapparent86} have highlighted the large-scale structure (hereafter LSS) of our Universe, a cosmic web formed by voids, sheets, elongated filaments and clusters at their nodes \citep{pogo96}. Characterizing quantitatively these striking features of the observed and modeled universe has proven to be both useful \citep{SDSSskel,gay} but challenging. It is useful because these features reflect the underlying dynamics of structure formation, and are therefore sensitive to the content   of the universe \citep{pogo}.  It is challenging because observations and simulations provide limited and noisy  data sets. Recently Soubie  (2010, hereafter paper I)  presented an algorithm able to estimate the underlying critical sets (walls, filaments, voids) from a given noisy discrete sample of the underlying field. Typically, this situation arises in astrophysics  when the aim is to recover the topology or the geometry  of the underlying density field while  only a catalogue of galaxies is available. For instance, in the context of understanding the history of our Milky Way,  it is of interest to identify the filaments of the local group. Yet typically in this context, only a limited number of galaxies at somewhat poorly estimated positions are observed. For redshift catalogues involving hundreds of thousands of galaxies, one would also wish to reconstruct the main features of the cosmic web as best as the non-uniform sampling allows.  From a theoretical point of view, it might for instance be of interest to compute the cosmic evolution of the filamentary network, as its history constrains the dark energy content of the universe. From an observational point of view, it could also help solving the missing baryon problem \citep{1998ApJ...503..518F} because most of such baryons has been considered to be located along the filamentary structure in the form of diffuse hot gas called Warm/Hot Intergalactic Medium  (WHIM; \cite{1999ApJ...514....1C}, \cite{aracil}).  Identifying the filament from galaxy distributions clearly provides good candidates for searching for the WHIM with UV absorptions \citep[e.g.]{2000ApJ...534L...1T,2010ApJ...710..613D}, X-ray absorptions \citep[e.g.][]{2002ApJ...564..604F,2006PASJ...58..657K,2009ApJ...695.1351B, 2010ApJ...714.1715F} and X-ray emission lines by future surveys \citep[e.g.][]{2003PASJ...55..879Y,2006SPIE.6266E..12O}. It is therefore of prime importance to provide a tool which deals consistently with such possibly sparse discrete samples.  Quite a few such options have been presented recently (\cite{skel2D,hahn,SDSSskel,rsex,skel,spine, aragonvoid,jaime,neyrinck,platen, stoica05,stoica07,stoica10}, \paperone), relying on different strategies on how to deal with these constraints (see \paperone).
 
 In paper I, the emphasis was on a formal presentation of the underlying mathematical theory and its extension to the discrete regime. As the corresponding algorithm is fairly 
intricate, a certain level of formal jargon was required to describe it unambiguously. 
Hence paper I focused on the language of mathematics. 
Conversely, let us first now rephrase here 
the corresponding framework in the more intuitive  language of astrophysical data processing, as our aim is to appeal to both the community of computational geometry and that of astrophysics. 
 What should be the expected characteristics of the ideal structure finder?
 Optimally, one would like to implement an algorithm which recovers the important and robust features of the underlying field even when little information is available, so that the procedure manages to reasonably identify the most striking features of the cosmic field.  {\sl Topolology} (in fact discrete topology) therefore provides  the  natural context in which the optimal algorithm should be implemented. 
 Indeed, topology de facto characterizes the  ``rubber" geometry of the underlying field, \ie its most intrinsic and robust features.
 More specifically, as argued in \paperone, 
 ideally such an optimal tool should produce an ensemble of critical sets (lines, surfaces  and volumes) consistent with those defined within the context of Morse theory for sufficiently smooth fields \citep{milnor,jost}.
 Morse theory basically provides a rigorous framework in which to formally define such sets for  ``regular" density fields (roughly speaking regular means not degenerate so that these sets are well defined). 
  For instance, the critical lines  defined by this theory connect peaks and 
 maxima  via special (extremal) flow lines of the gradient\footnote{Indeed, Morse theory formalizes the process of partitionning space according to the gradient flow of the density into so called ascending and descending manifolds. In other words, it tags space according to where one would end up going ``uphill" or ``downhill". In doing so it identifies  special lines or surfaces, where something unusual happens.}. These lines should trace quite well the filaments of the LSS.  Similarly, the walls of the LSS  should have 
 a natural equivalent feature as the ``critical" surfaces of Morse theory (the so called 2-manifolds).
 
    Here our purpose is to  proceed within the context of its discrete counterpart \citep{forman}.
The corresponding  discrete construction should be  as consistent as possible with all the topological features of  the underlying smooth field  (it should globally preserve, at the level  of the noise, the salient  features of the field, such as the number of connected components, the number of tunnels or holes defined by its iso-contours, etc.; conversely\footnote{this well known duality between the topology of the level sets and  the characteristics of the critical points clearly has a discrete analogue through the creation/destruction of discrete cycles, see paper I} the significant discrete critical sets should have the correct ``combinatorial" properties: \eg critical lines should only connect  at critical points, saddle points  connect exactly two peaks together, etc...). The level of complexity of the corresponding network should also reflect the inhomogeneities of the underlying survey: \ie it should adapt its level of description to the sampling,  hence yielding a parameter-free multiscale description of the cosmic web. In fact, it should also provide a simple diagnostic in order to estimate the robustness of the various components of the network (\ie the degree of reproduced details should be tunable to the purpose of the investigation). Finally it should clearly address the shortcomings of watershed-based methods described in paper I 
(namely the occurrence of spurious boundary lines induced by resampling in 3D or more).

 Paper I presented precisely such a tool, while building up on an extension of Morse theory to discrete unstructured meshes. 
 Two main challenges were addressed: (i) defining the counterpart of the (topologically consistent) critical sets {\sl on} the mesh; (ii) defining a procedure to simplify the corresponding mesh
 at the level of the local shot noise. \\ The first step is achieved by considering simultaneously all the discrete components of the triangulated mesh (its vertices, edges, faces and tetrahedra),
 and reassigning a density to all these components in a manner which is heuristically consistent with the sampled density at the vertices; this relabeling procedure also ensures that the discrete flow (which follows from the corresponding discrete gradient) is sufficiently well-behaved to provide such topological consistency (specifically, it ensures the  existence of  
 discrete counterparts of regular critical points). Loosely speaking, amongst the discrete analogues of gradient flows, the algorithm identifies the critical subsets as  special sets which cannot be  paired to their neighbours through these discrete gradients.
 Note that the required level of compliancy  to achieve this construction has the virtue of not only producing the discrete set of critical segments, but also the triangulated walls and voids at no extra computational  cost.  
 \\ The second step follows from the concept of topological persistence \citep{edel00,edel}, which  assigns a 
 density ratio to pairs of critical points which are found to be connected together by such discrete integral paths; { these pairs are identified  by the destruction/creation of  critical points as one describes the level sets}.
  If this ratio is below a given threshold, the corresponding critical line/surface is found to be (topologically) insignificant,  it is removed from the set and the remaining critical mesh is simplified so as to recover some topological consistency. In other words,
 if the shot noise induces the occurrence of the discrete counterparts of, \eg spurious loops, disconnected blobs, or tunnels which are found to be insignificant according the the aforementioned criterium, they will be removed. The idea of topological persistence is central in producing a natural (topologically motivated)
 self-consistent criterion for infering the significance level of the identified structures. {In particular, it warrants that the removal of pairs of critical points  consistently extracts the corresponding topological feature (loop, tunnel, component).}
 
Importantly, let us emphasize that within this framework, the mathematical theories that we use are intrinsically discrete and readily apply to the measured raw data (modulo the consistent labelling of the elements of the Delaunay tessellation  relative to the DTFE densities computed at the sampling points). This warrants that all the well-known and extensively studied mathematical properties of Morse theory are ensured by construction {\sl at the mesh level}, and that the corresponding cosmological structures therefore correspond to well-defined mathematical objects with known mathematical properties.

In the language of computational geometry (see Appendix \ref{sec_terminology} for the relevant definitions), a simplicial complex (the tessellation) is computed from a discrete distribution (galaxy catalogue, N-body simulation, ...) using  a Delaunay tessellation. A density $\rho$ is assigned  to each galaxy using DTFE (roughly speaking, the density at a vertex is proportional to the inverse volume of its dual Voronoi cell, see \citet{DTFE}). A discrete Morse function (a re-labelling of all elements of the tessellation) is then defined by attributing a properly chosen value to each simplex in the complex (\ie the segments, facets and tetrahedron of the tessellation). From this discrete function, we then compute the \hyperref[defDG]{discrete gradient} and deduce the corresponding \hyperref[defDMC]{Discrete Morse-Smale complex} (DMC hereafter, \citet{forman}). The \hyperref[defDMC]{DMC} (the set of critical points connected by \hyperref[defarc]{arcs}, quads, \hyperref[defcrystal]{Crystals} etc..) is  used as the link between the topological and geometrical properties of the density field. Its critical points together with their ascending and descending \hyperref[defmanifold]{manifolds} (the ``critical" sets) are identified to the peaks, filaments, walls and voids of the density field. The \hyperref[defDMC]{DMC} is then filtered  using persistence theory. For that purpose, we consider the \hyperref[deffiltration]{Filtration} (the discrete counter part of the density-sorted level-sets) of the tessellation according to the values of the discrete Morse function and use it to compute persistence pairs of critical points (pairs of critical points which are linked by their appearance and disappearance as one scans the \hyperref[deffiltration]{Filtration}). The \hyperref[defDMC]{DMC} is simplified by canceling the pairs that are likely to be generated by noise. This is achieved by computing the probability distribution function of the persistence ratio (\ie the ratio of the densities {\sl at} the connected pair) of all types of pairs in scale-invariant Gaussian random fields and canceling the pairs with a persistence ratio whose probability is lower than a certain level.

Paper I presented a couple of examples of such a construction in 2D. Let us now illustrate the virtue of the method in the context of the 3D cosmic Web. 
We start\footnote{Note that our goal here is not to present an exhaustive review of the geometrical properties of the cosmic web, which is clearly out of the scope of this paper.} by  showing that the geometry of the cosmic web is accurately reproduced, while 
illustrating the quality of the cosmological structures identification, both in an N-Body simulation
(section~\ref{sec:LSSsimu}) and directly on an unprocessed version of the SDSS DR7 galaxy catalogue (section~\ref{sec_SDSS}). 
In particular we show how  \progname\, allows us to identify  various configurations of the filamentary structure of galaxies,
 and identify a previously missed X-ray ``optically faint" halo  at the intersection of a set of SDSS filaments using the SUZAKU satellite. 
 We then discuss in section~\ref{subsec_sigthres}  the problem of estimating the right value for the persistence  level in cosmological  simulations,
  and illustrate how the measured topological properties of the LSS distributions are affected by varying this threshold. 
 In particular we show how this criterion compares with the simple friend-of-friend algorithm when attempting to identify 
halos in simulations. Section~\ref{sec_conclusion} wraps up and discusses prospects.

\section{Geometry of LSS: simulation}
\label{sec:LSSsimu}

\begin{figure*}
\begin{minipage}[c][\textheight]{\linewidth}
\begin{minipage}[c]{0.49\linewidth}
\centering\includegraphics[height=0.3\textheight]{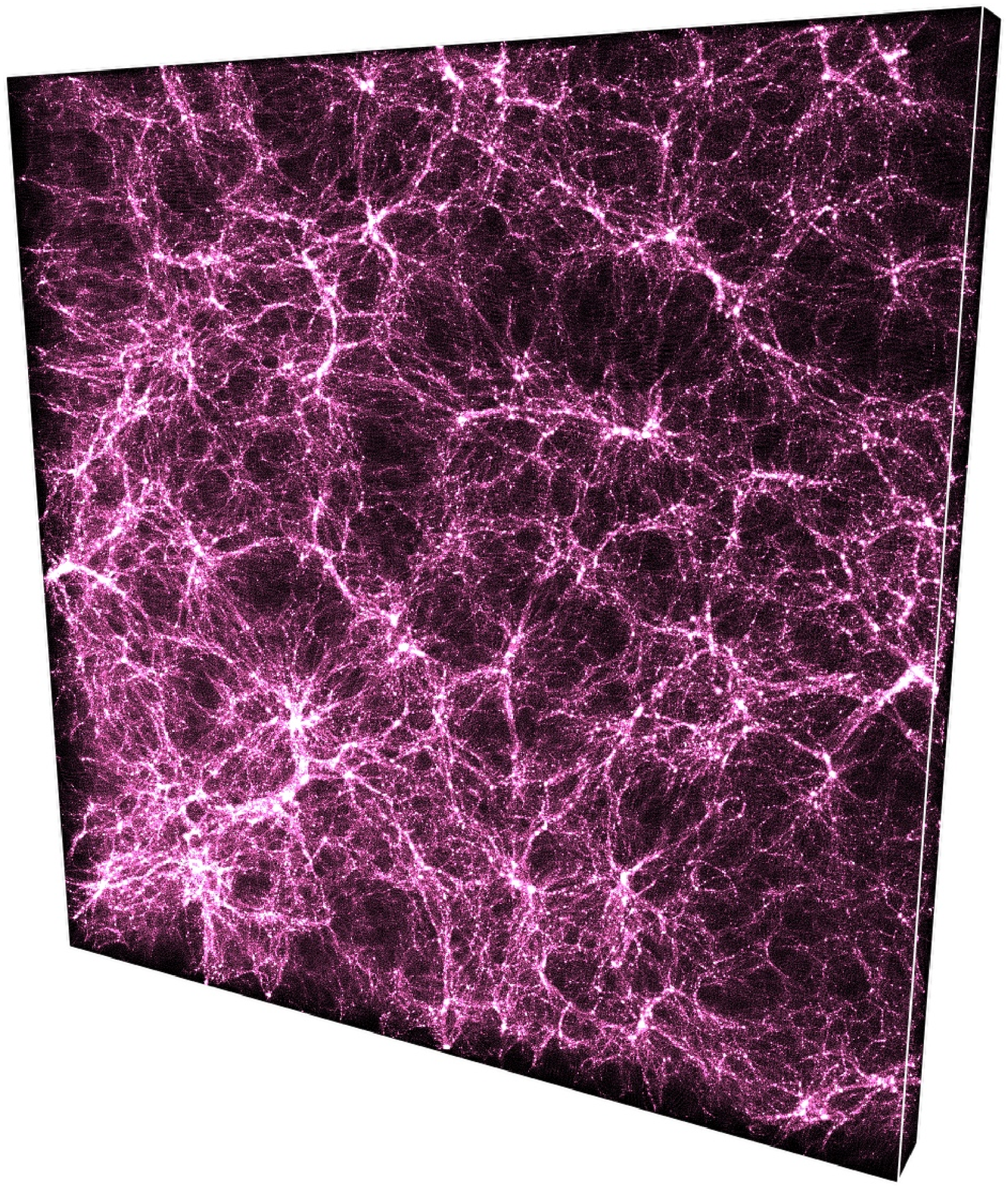}\\
\centering\includegraphics[height=0.3\textheight]{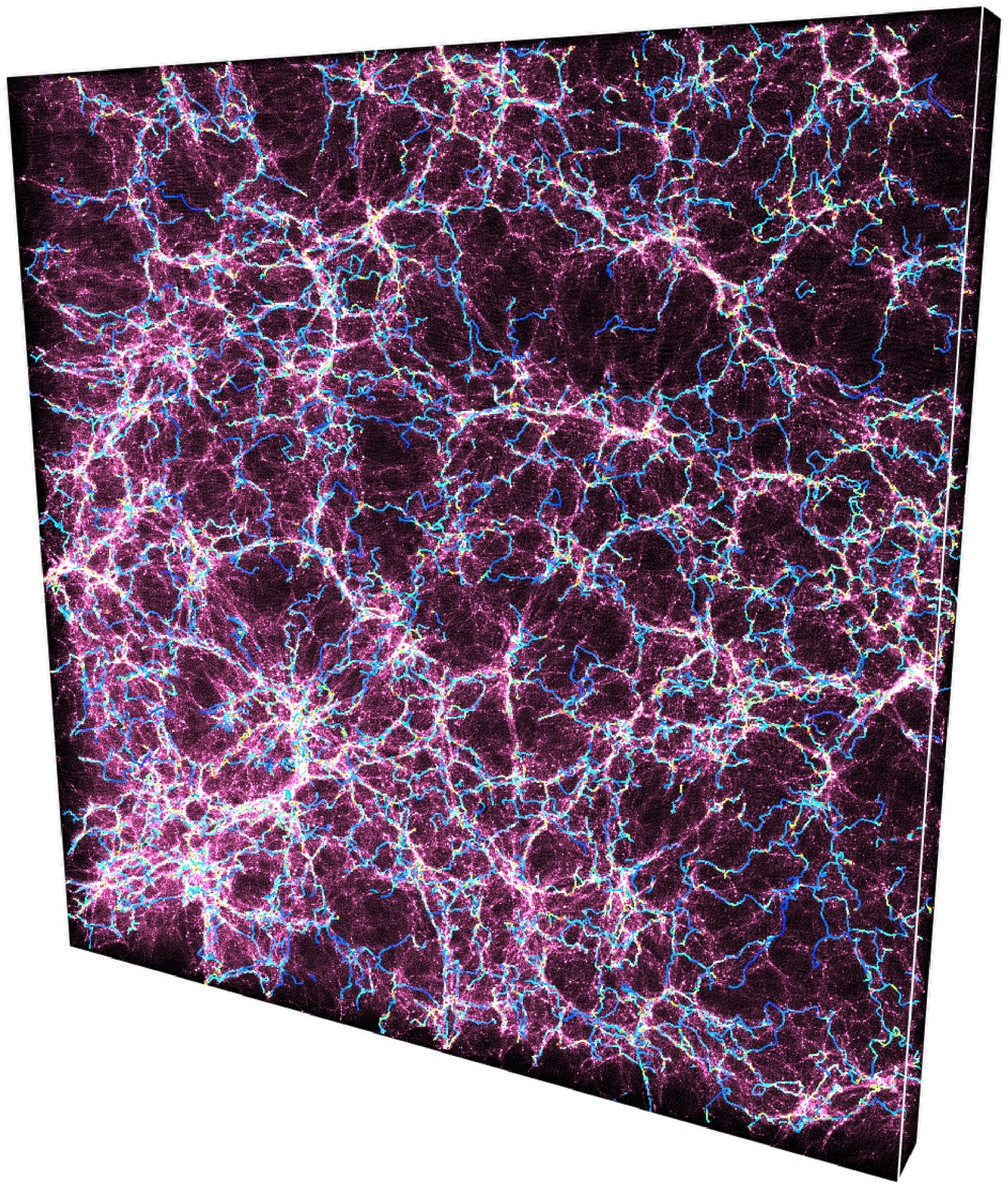}\\
\centering\includegraphics[height=0.3\textheight]{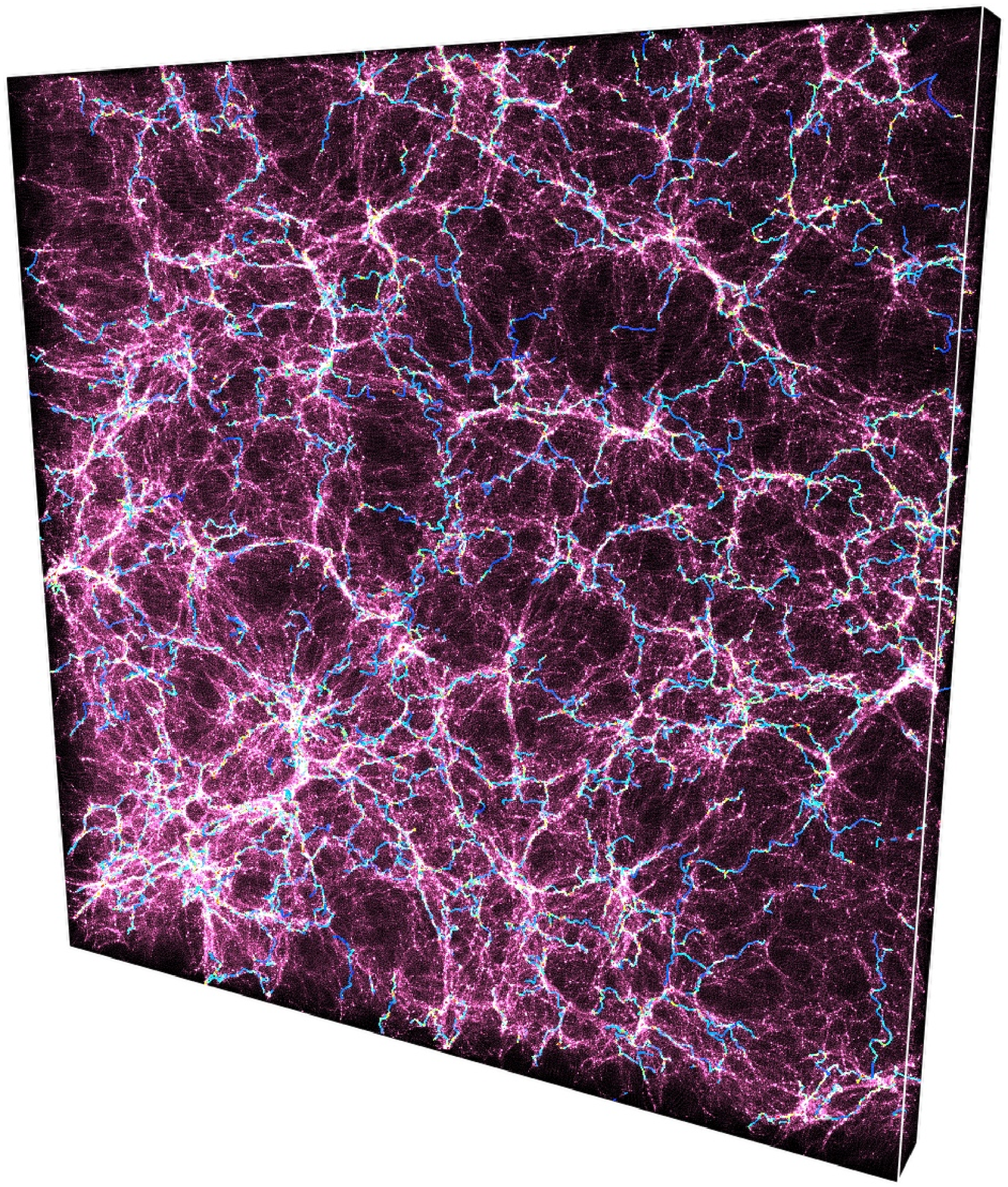}\\
\end{minipage}
\hfill
\begin{minipage}[c]{0.49\linewidth}
\centering\includegraphics[height=0.3\textheight]{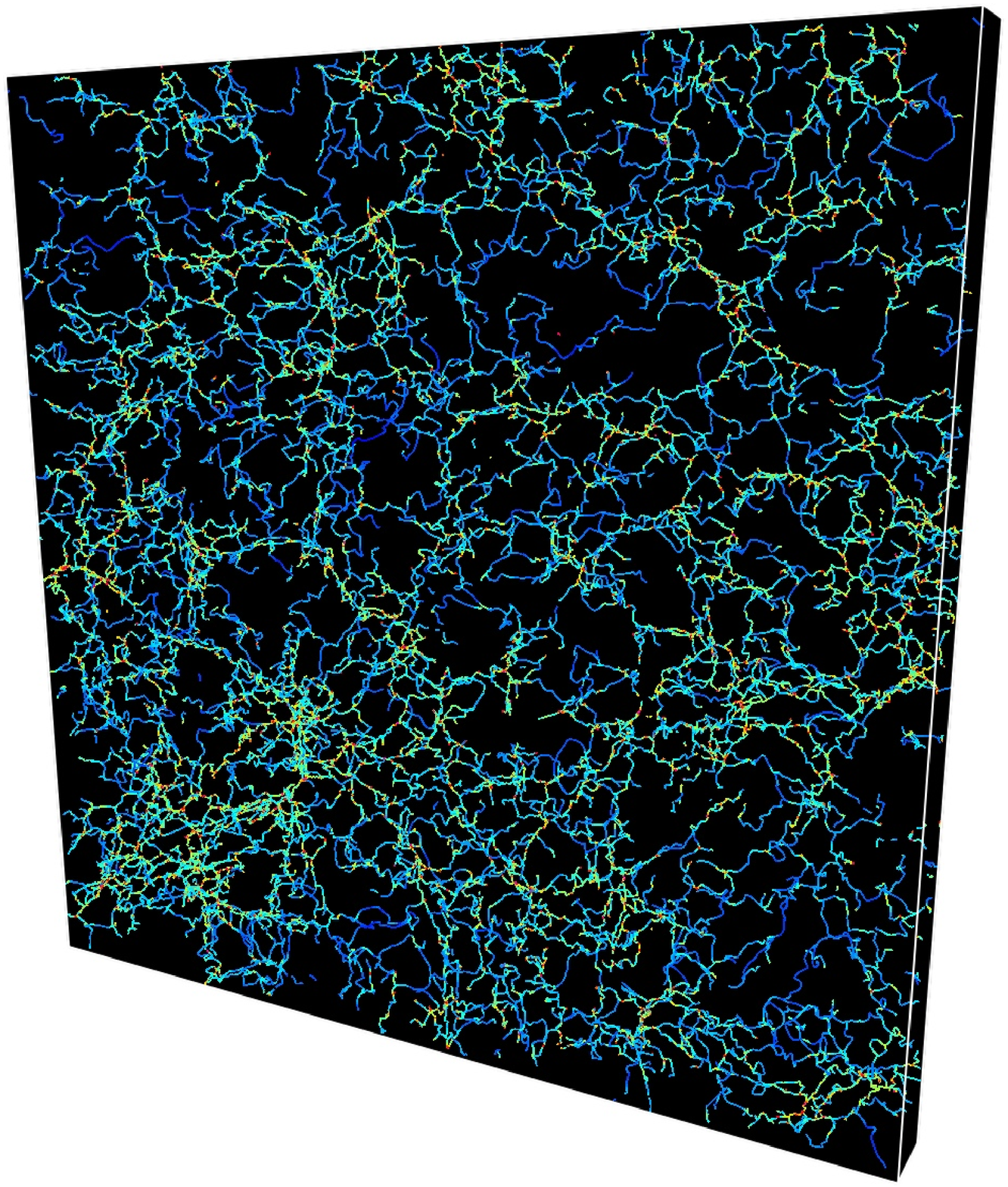}\\
\centering\includegraphics[height=0.3\textheight]{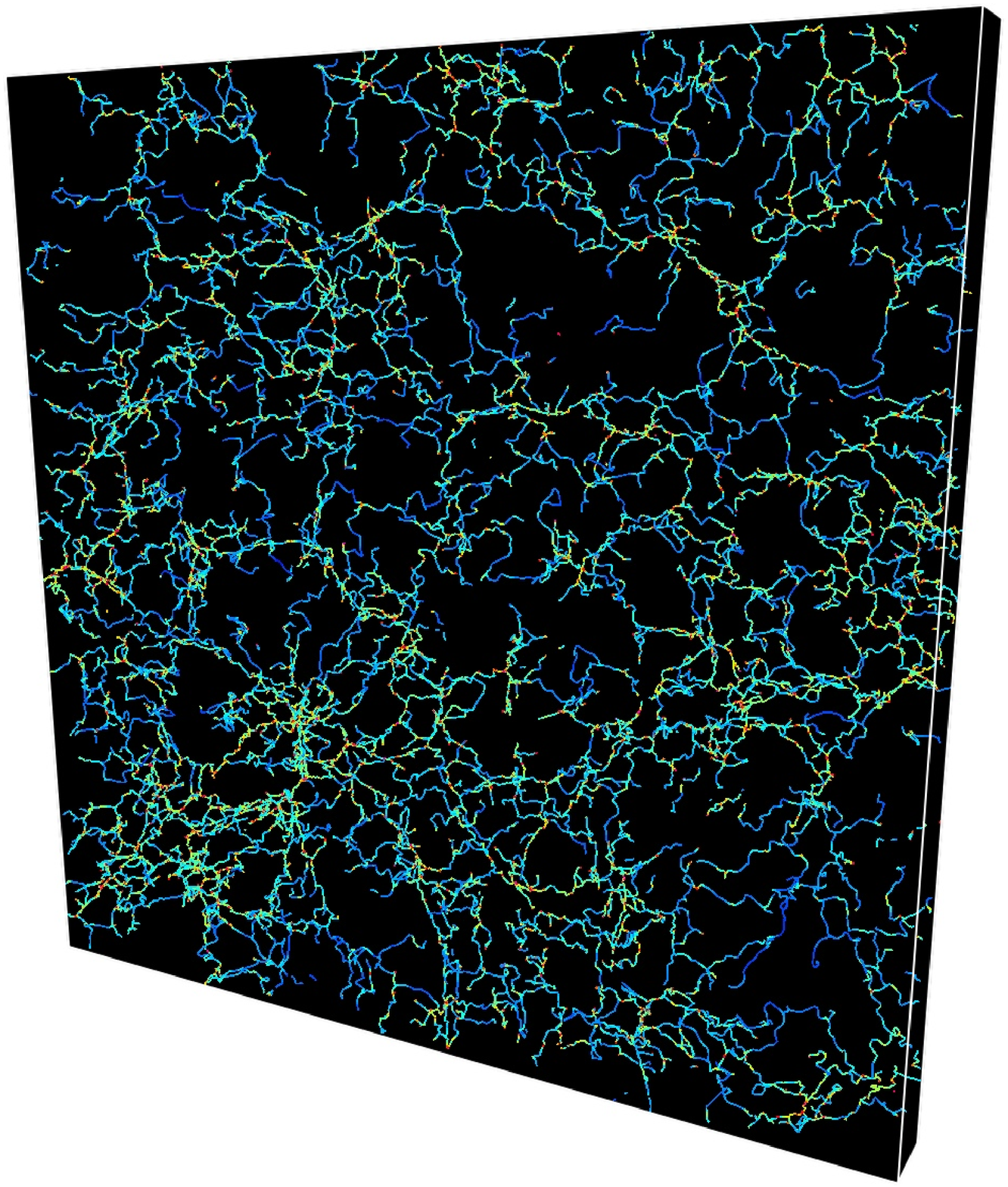}\\
\centering\includegraphics[height=0.3\textheight]{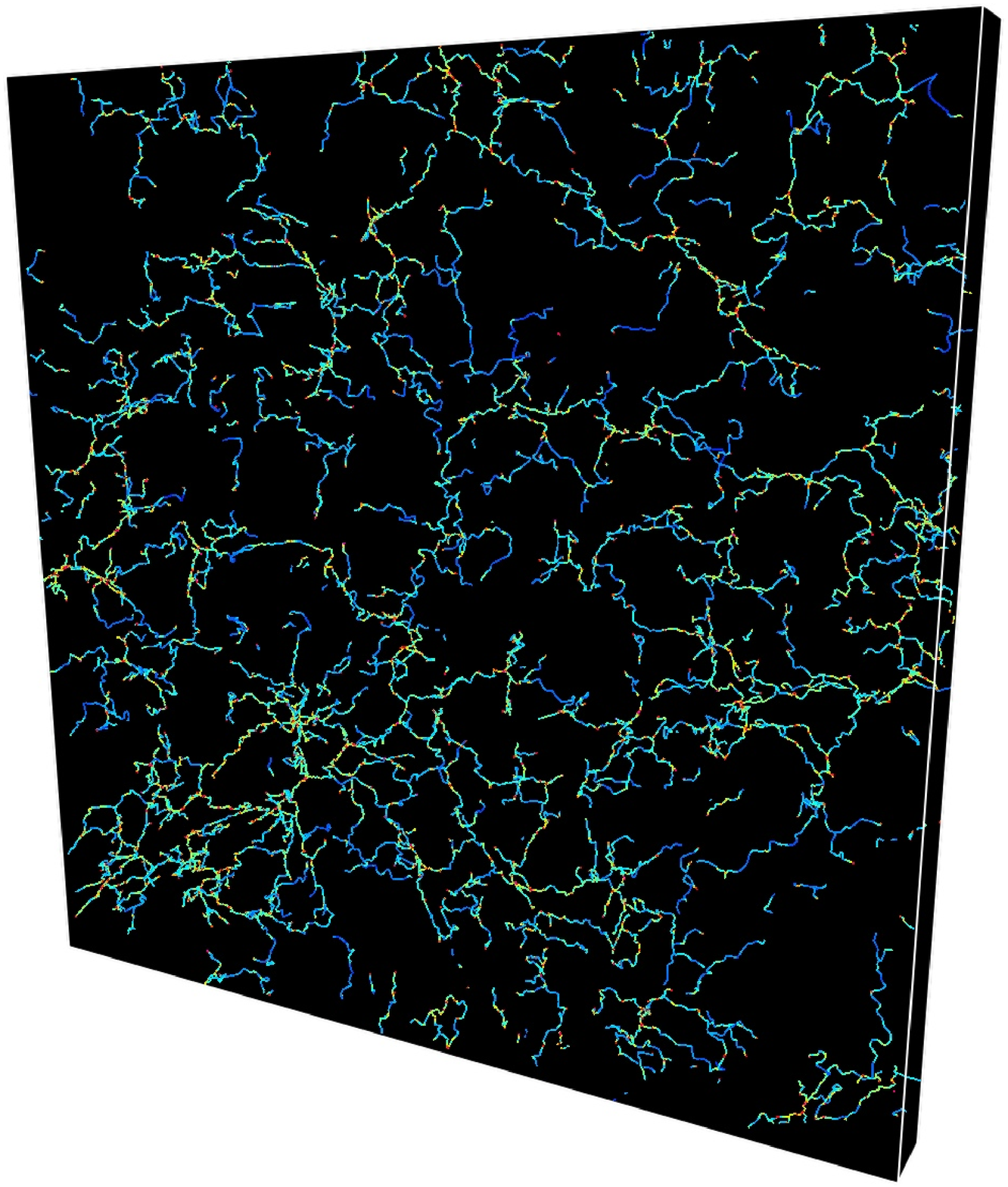}\\
\end{minipage}
\caption{The filamentary distribution above a persistence level of $\nsig{3}$, $\nsig{4}$ and $\nsig{5}$ (from top to bottom) in a $250\times250\times20\Mpc$ slice of a $512^3$ particles and $250\Mpccc$ large cosmological simulation. The computation was achieved on a $128^3$ particles sub-sample, and the filaments are colored according to the logarithm of the density. The density field was represented using the $512^3$ particles of the N-body simulation. \label{fig_simu250} }
\end{minipage}
\end{figure*}

\begin{figure*}
\begin{minipage}[c][\textheight]{\linewidth}
\centering\subfigure[Simulated dark matter distibution]{\includegraphics[width=0.49\linewidth]{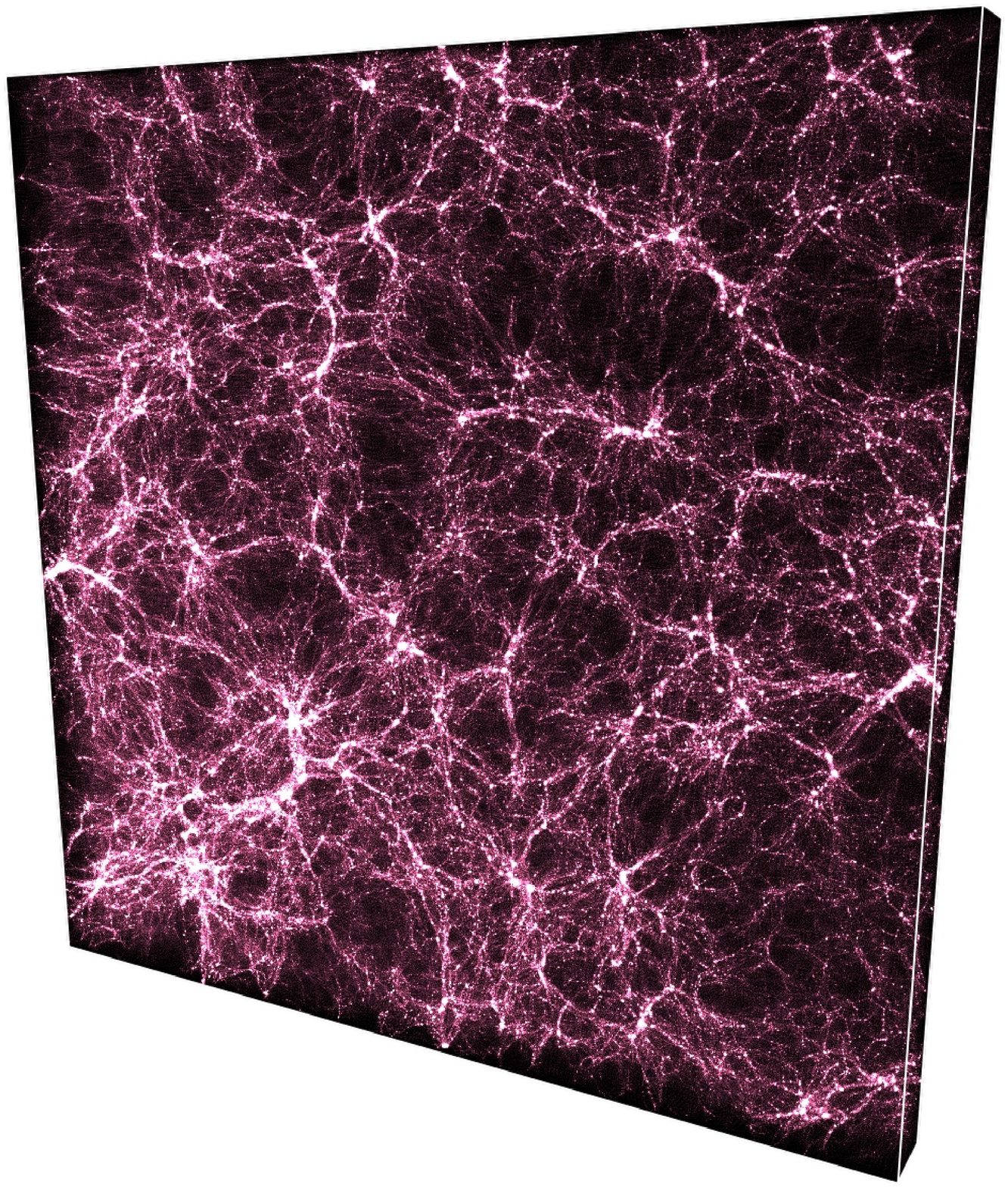}\label{fig_simu250_voidA}}
\hfill\subfigure[A void (bottom right) embedded in the filamentary structure]{\includegraphics[width=0.49\linewidth]{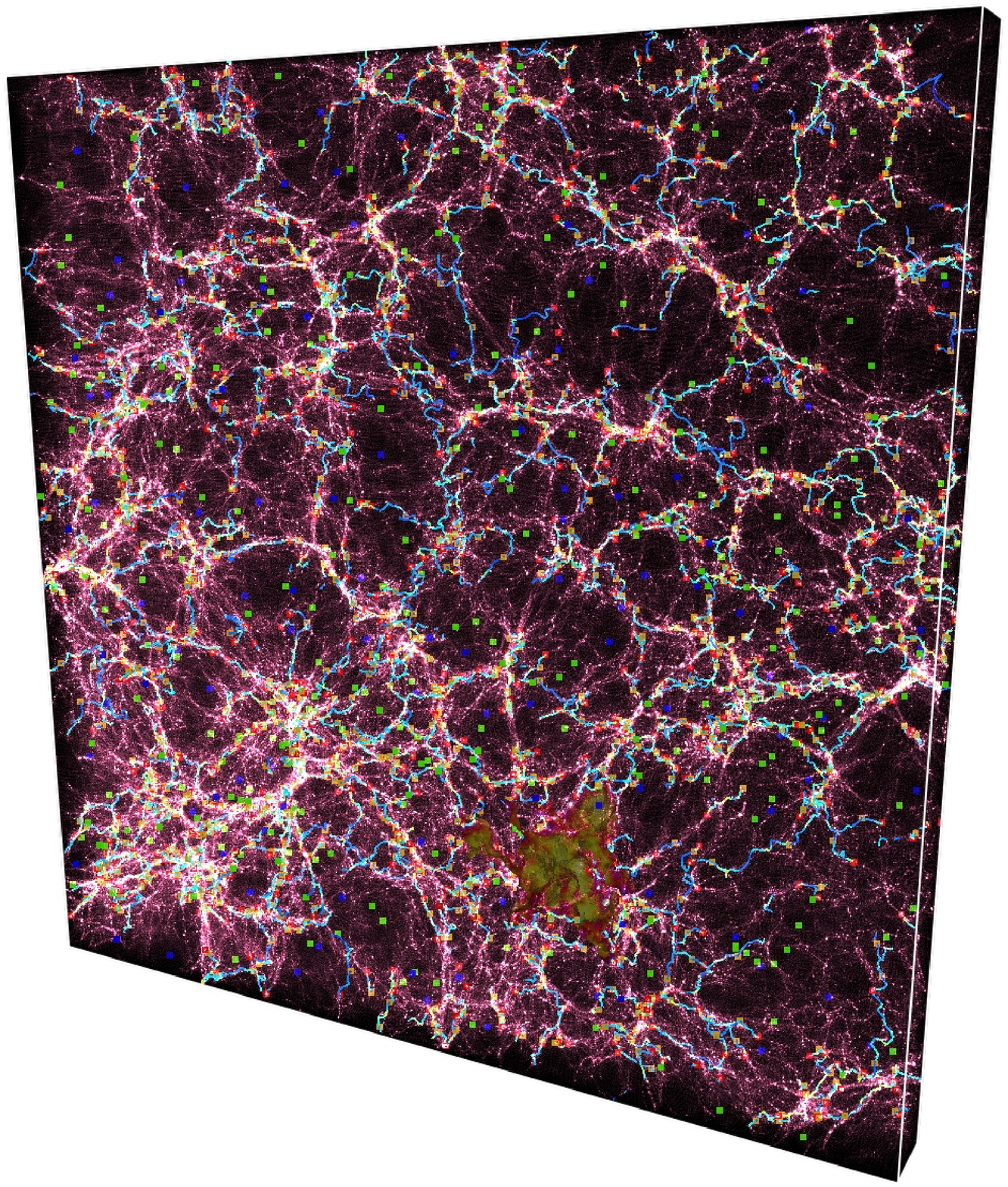}\label{fig_simu250_voidB}}\\
\centering\subfigure[Zoom on the void of panel \ref{fig_simu250_voidB}.]{\includegraphics[width=0.49\linewidth]{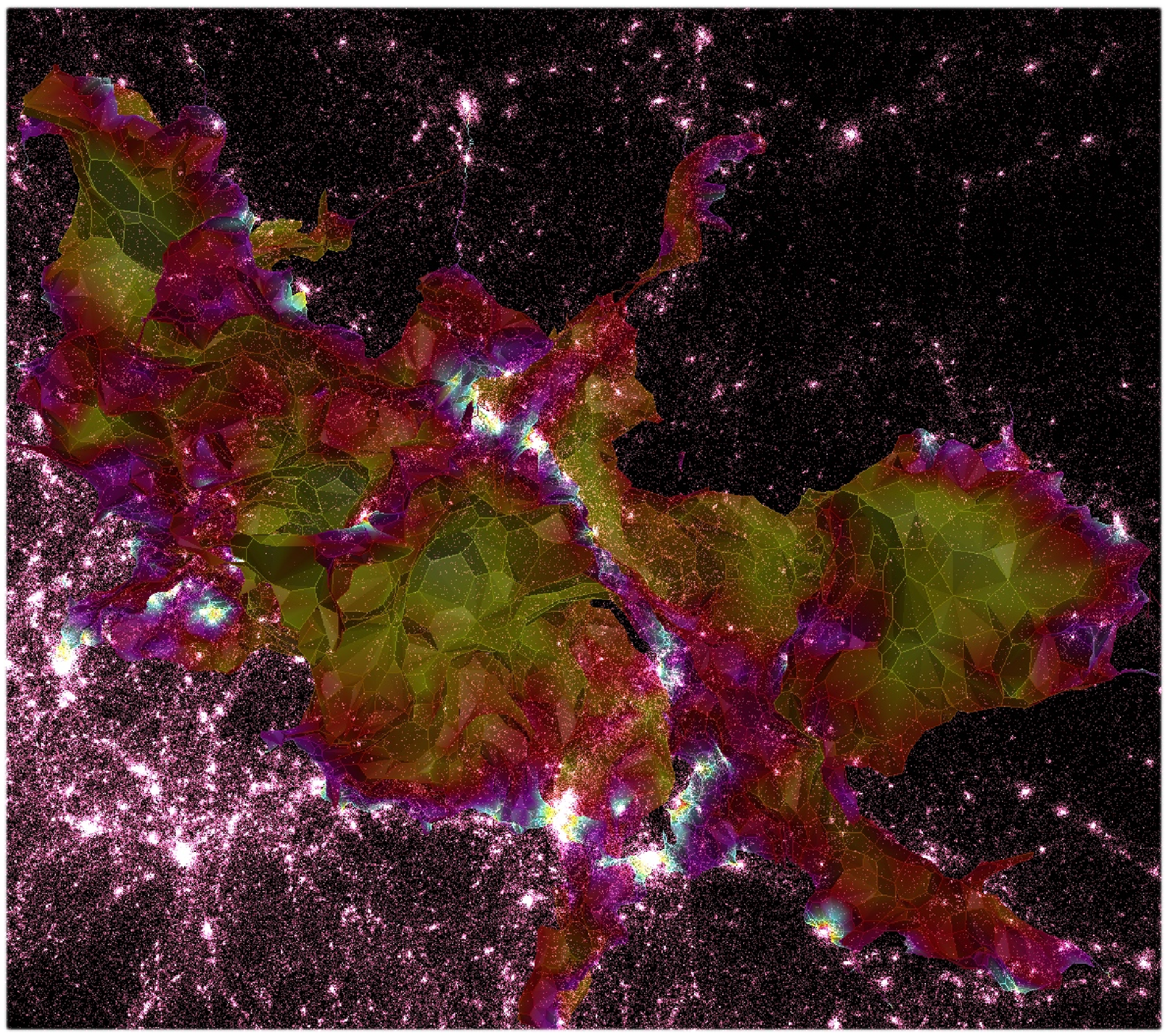}\label{fig_simu250_voidC}}
\hfill\subfigure[The relationship between the detected void, filaments and critical points]{\includegraphics[width=0.49\linewidth]{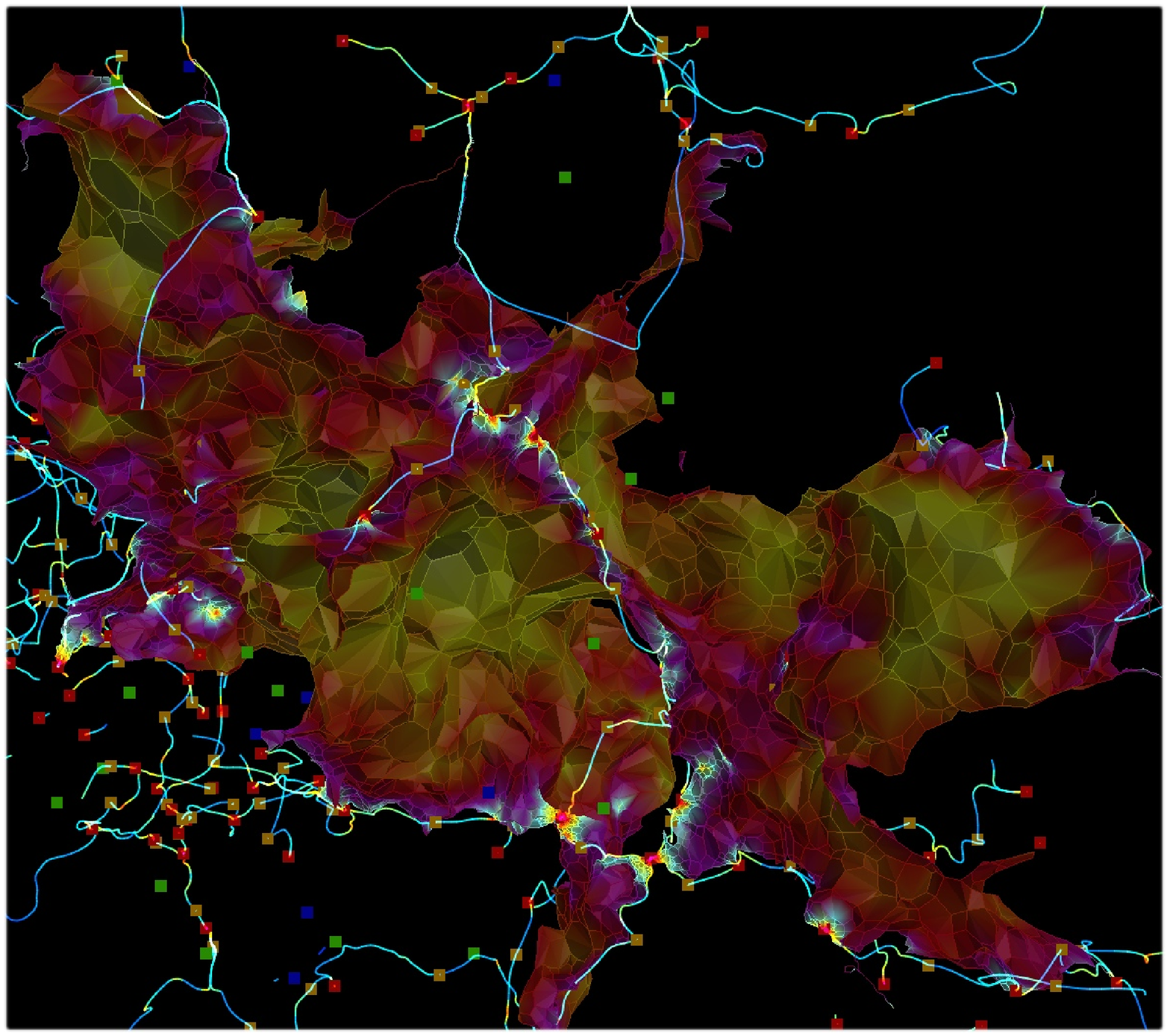}\label{fig_simu250_voidD}}\\
\caption{The \hyperref[defarc]{arcs} of the \hyperref[defDMC]{Discrete Morse-Smale complex} (\ie the filaments) and an ascending $3$-\hyperref[defmanifold]{manifold} (\ie a void) at a significance level of $\nsig{5}$ in the same distribution as that of figure \ref{fig_simu250} (a $250\times250\times20\Mpc$ slice of a $512^3$ particles and $250\Mpccc$ large cosmological simulation). The density distribution is represented using all available particles within the simulation (panel \ref{fig_simu250_voidA}) while the DMC was computed using $128^3$ particles sub-sample. The $2$ lower panels (\ref{fig_simu250_voidC} and \ref{fig_simu250_voidD}) show a zoom on the upper panels at the location of the randomly selected void (see panel \ref{fig_simu250_voidB}). On these figures, the maxima, $1$-saddle points, $2$-saddle points and minima are represented as red, yellow, green and blue square respectively and the \hyperref[defarc]{arcs} as well as the manifold are shaded according to the log of the density. Note on panel \ref{fig_simu250_voidD} how the maxima, saddle-points and path of the filaments corresponds to the crests of the 2D density field measured on the surface of the void. This particularly emphasize the coherence of the detection of objects of different nature. \label{fig_simu250_void}}
\end{minipage}
\end{figure*}

Although we have shown in paper I that the method introduced in that paper seems to be able to measure the topological properties of the cosmic web efficiently even in the presence of significant noise, we only illustrated in the 2D case that it could also recover correctly the geometry of the filamentary structure (see paper I). Demonstrating that a given algorithm is able to correctly identify the location of filaments is a difficult task,  as it  requires the previous knowledge of the location of those filaments. The only solution therefore seems to  build an artificial distribution from a previously defined filamentary structure. This method was adopted in \citet{spine}, where the authors use a Voronoi Kinematic Model \citep{weygaert02}. The principle of the Voronoi Kinematic Model is to identify the voids, walls, filaments and clusters to the cells, faces, edges and vertice of the Voronoi tesselation. In practice, randomly distributed particles are moved away from the nuclei of the Voronoi cells following a universal expansion rate, and their displacement being constrained to the faces, edges and finally vertice as they reach them. This results in a distribution of particles where each is said to be a void, wall, filament or cluster particle depending on weather they belong to a cell, face, edge or vertice of a Voronoi cell when the simulation is stopped.\\

We argue that using such a model to quantify the quality of the Morse-Smale complex identification is not as relevant as one would think, mainly because it is too idealized topologically speaking. In fact, it is a built-in property of the Voronoi Kinematic Models that all the cosmological structures overlap neatly: maxima (\ie voronoi vertice) are located at the intersection of filaments (\ie Voronoi segments) that always intersect with a suitable angle, those filaments are themselves by definition located at the intersection of at least three voids (\ie voronoi cells), and each pair of neighboring void have exactly one Voronoi face in common, neatly defining the walls. As was shown in paper I, density functions extracted from actual data sets are in fact quite different, as they do not comply so easily to Morse conditions, in particular when measured from cosmological simulations or observational galaxy catalogues. In that case, and as clearly shown in paper I (see appendix 1), filaments may (and actually often do) merge before reaching a maximum, two apparently neighboring voids (down to the resolution limits) do not necessarily share a common face, and filaments are not necessarily at the intersection of at least three voids (once again, down to the resolution limit). The nature of the  Voronoi Kinematic Model is therefore such that it  avoids all the difficulty in identifying the Morse-Smale complex of realistic data sets. It might be possible to build more sophisticated Voronoi Models, that would for instance mimic the structure mergers that occur along the course of the evolution of large scale matter distribution in the Universe, but this is clear out of the scope of this paper. For the lack of a simple better way, we therefore use here what is probably to date the most efficient way to detect structures: the human eye and brain.\\

\begin{figure*}
\begin{minipage}[c][\textheight]{\linewidth}
\centering\subfigure[Dark matter distribution in a $50\times50\times20\Mpc$ sub-box]{\includegraphics[width=0.49\linewidth]{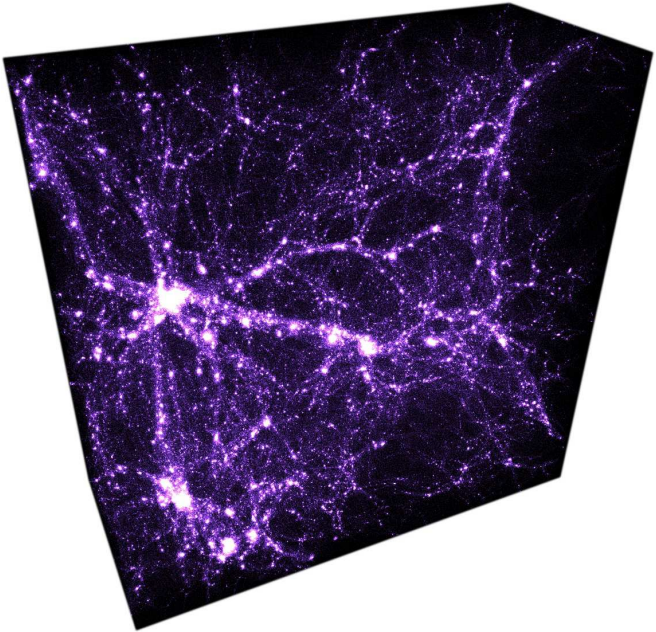}\label{fig_simu_manifolds_picA}}
\hfill  \subfigure[An ascending $2$-manifold (\ie a wall)]{\includegraphics[width=0.49\linewidth]{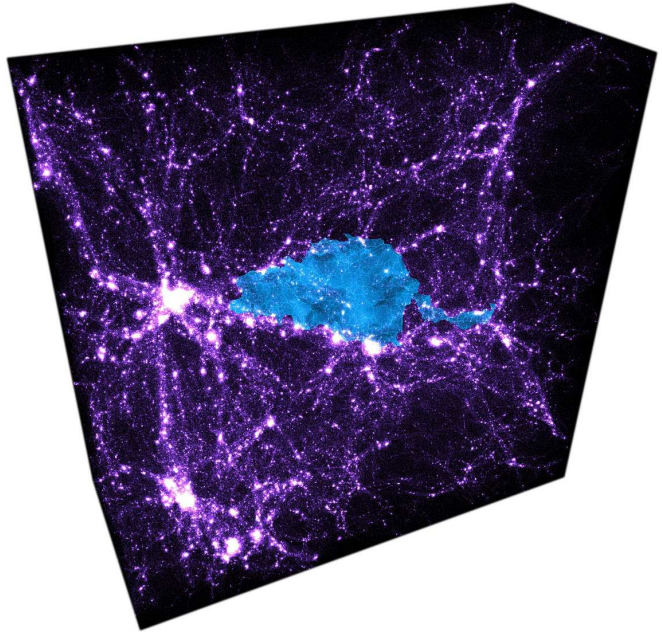}\label{fig_simu_manifolds_picB}}\\
\centering\subfigure[An ascending $3$-manifold (\ie a void)]{\includegraphics[width=0.49\linewidth]{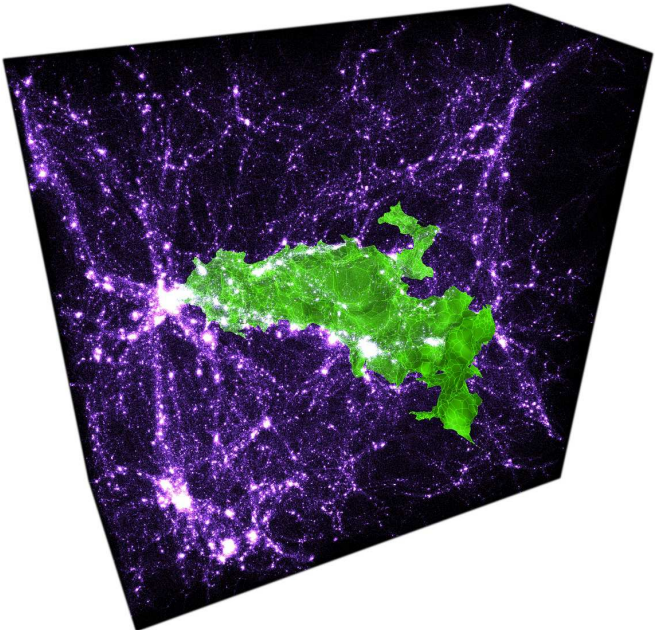}\label{fig_simu_manifolds_picC}}
\hfill  \subfigure[Superposition of an ascending $3$-manifold and an ascending $2$-manifold on its surface.]{\includegraphics[width=0.49\linewidth]{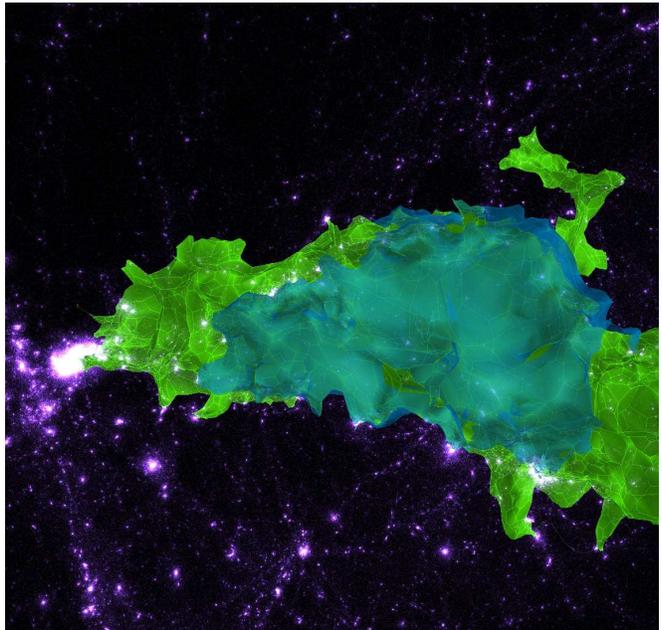}\label{fig_simu_manifolds_picD}}\\
\caption{An ascending $2$-manifold (\ie blue 2D wall) and an ascending $3$-manifold (\ie green 3D void) identified in a $512^3$ particles $100\Mpc$ $\lcdm$ dark matter simulation. The manifolds where computed from a $64^3$ particles sub-sample.\label{fig_simu_manifolds_pic}}
\end{minipage}
\end{figure*}

\subsection{Visual inspection}

The evolution of the geometry of the measured filaments with significance threshold is illustrated on figure \ref{fig_simu250}. The \hyperref[defDMC]{DMC} represented on this figure was computed at significance levels of $3$, $4$ and $\nsig{5}$ (from top to bottom) within $128^3$ particles sub sample of a $512^3$ particles, $250\Mpccc$ $\lcdm$ dark matter only N-body simulation. Note that the dark matter distribution within the $20\Mpc$ slice is represented in the top left corner to facilitate the visualization of its filamentary structure. Despite the projection effects that create visual artifacts (\ie spurious filament looking structures resulting from the projection of dark matter clumps at different depths) and the fact that filaments may enter or leave the slice, therefore seemingly appearing and disappearing for no apparent reasons, it seems fair to recognize that the agreement between the observed and measured filaments is excellent. These good performances are mainly the result of our use of the scale adaptive Delaunay tessellation and the fact that our implementation does not require any pre-treatement of the density field, unlike usual grid based methods which enforce a maximal resolution and resort to some kind of density smoothing technique that affect the geometrical properties of the distribution. As a result, the resolution of the filaments is optimal with respect to the initial sampling whatever the selected significance level: the higher \hyperref[defpers]{persistence} and larger scale filamentary network is, by construction, a subset of its less persistent and lower scale counterpart. Because \hyperref[defpers]{persistence} based topological simplification is used, increasing the \hyperref[defpers]{persistence} threshold actually results in less significant filaments disappearing (when simplifying a $1$-saddle point/$2$-saddle point \hyperref[defpers]{persistence} pair) or merging into each other (when simplifying a $1$-saddle point/maximum \hyperref[defpers]{persistence} pair) to form larger scale more persistent ones, but conserving the exact same resolution in any case. This can easily be observed by comparing the filamentary networks on the right column of figure \ref{fig_simu250}.\\

Another remarkable advantage of constructing cosmological structures identification on Morse theory is the extraordinary built-in coherence of the results, whatever the type of structure, as shown on figures \ref{fig_simu250_void} and \ref{fig_simu_manifolds_pic}. For instance, the intricate pattern of a randomly selected void (\ie an ascending $3$ \hyperref[defmanifold]{manifold}) embedded within the filamentary network (\ie ascending $1$ \hyperref[defmanifold]{manifolds}) of the cosmic web in the same simulation as previously is shown on figure \ref{fig_simu250_void}. The location of the void within the slice is displayed on panel \ref{fig_simu250_voidB}, each colored square standing for a critical point (see legend). On the zoomed frame (\ref{fig_simu250_voidC} and \ref{fig_simu250_voidD}), the surface of the void has been shaded according to the logarithm of the density, showing how the \hyperref[defDMC]{DMC} correctly traces the filamentary structure at the interface of the ascending $3$-\hyperref[defmanifold]{manifolds}, as expected in Morse theory\footnote{The slight shift in position between the surface of the void and the filament is due to the fact that we smoothed the filaments $4$ times (see paper I)}. Similarly, the neat relationship between a detected void and a wall structure on its surface (\ie an ascending $2$ \hyperref[defmanifold]{manifold}) in a $100\Mpc$ large N-body simulation is displayed on figure \ref{fig_simu_manifolds_pic}.\\

\begin{figure}
\centering
\includegraphics[width=\linewidth]{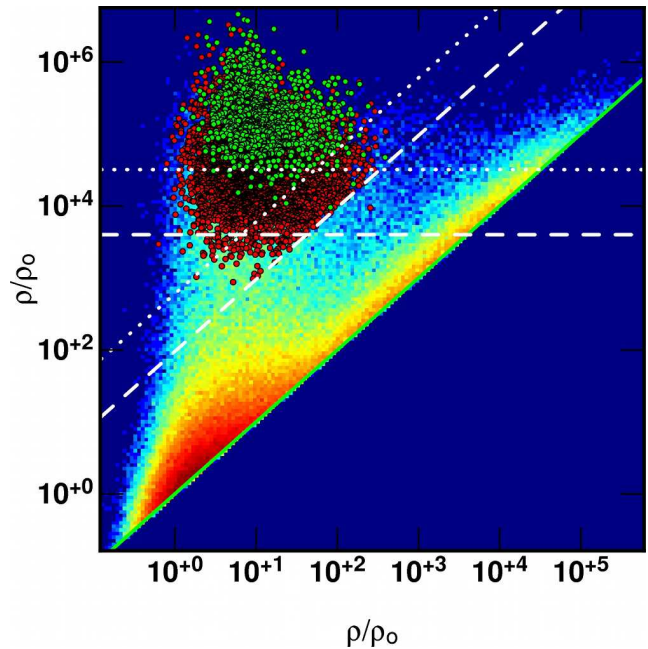}
\caption{Distribution of the \hyperref[defpers]{persistence} pairs of the highest density particles within each dark matter halo of mass $M>74 \times 10^{10}\;M_\odot$ (red) and $M>590 \times 10^{10}\;M_\odot$ (green) in a $128^3$ particle sub-sample of a $100\Mpc$ large $\lcdm$ dark matter simulation. The \hyperref[defpers]{persistence} diagram of maxima/$1$-saddle-points pairs with \hyperref[defpers]{persistence} larger than $\nsig{3}$ is shown in the background. The horizontal dashed and dotted lines correspond to overdensity levels of $4\times 10^3$ and $3.2\times 10^4$ respectively and the oblique lines to \hyperref[defpers]{persistence} levels of $\sim\nsig{4}$ and $\sim\nsig{5}$ respectively.\label{fig_halo_per_diag}}
\end{figure}

\begin{figure}
\centering\subfigure[Dark matter distribution in a $50\times50\times20\Mpc$ sub-box and haloes with mass $M>73.8\;10^{10}\;M_\odot$]{\includegraphics[height=0.26\textheight]{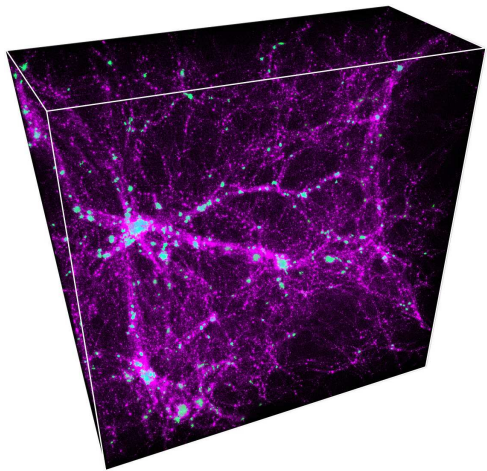}\label{fig_simu_halos_picA}}\\
\subfigure[The filaments at $\nsig{4}$ on a $128^3$ sub-sample]{\includegraphics[height=0.26\textheight]{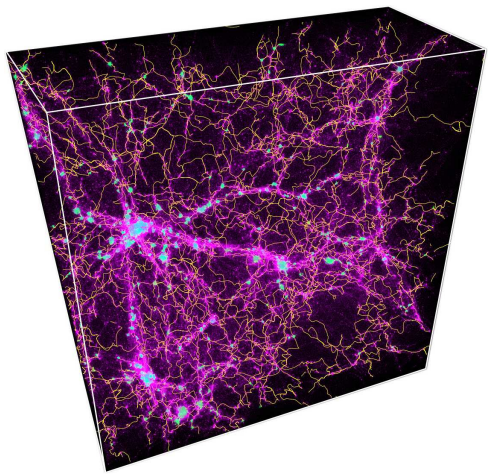}\label{fig_simu_halos_picB}}\\
\subfigure[The main filaments of the dark matter haloes]{\includegraphics[height=0.26\textheight]{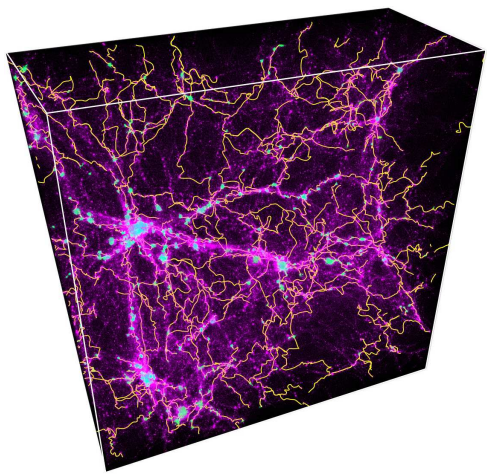}\label{fig_simu_halos_picC}}\\
\caption{Distribution of the main filaments of FOF haloes with mass $M>73.8\;10^{10}\;M_\odot$ in a $20\Mpc$ thick slice of a $512^3$ particles $100\Mpccc$ $\lcdm$ dark matter simulation. The filaments were computed from a $128^3$ particles sub-sample. Note that many filaments are linked to halos outside the slice, giving the false impression to end for no reason.\label{fig_simu_halos_pic} }
\end{figure}

\begin{figure}
\begin{minipage}[c][\textheight]{\linewidth}
\centering\includegraphics[height=0.28\textheight]{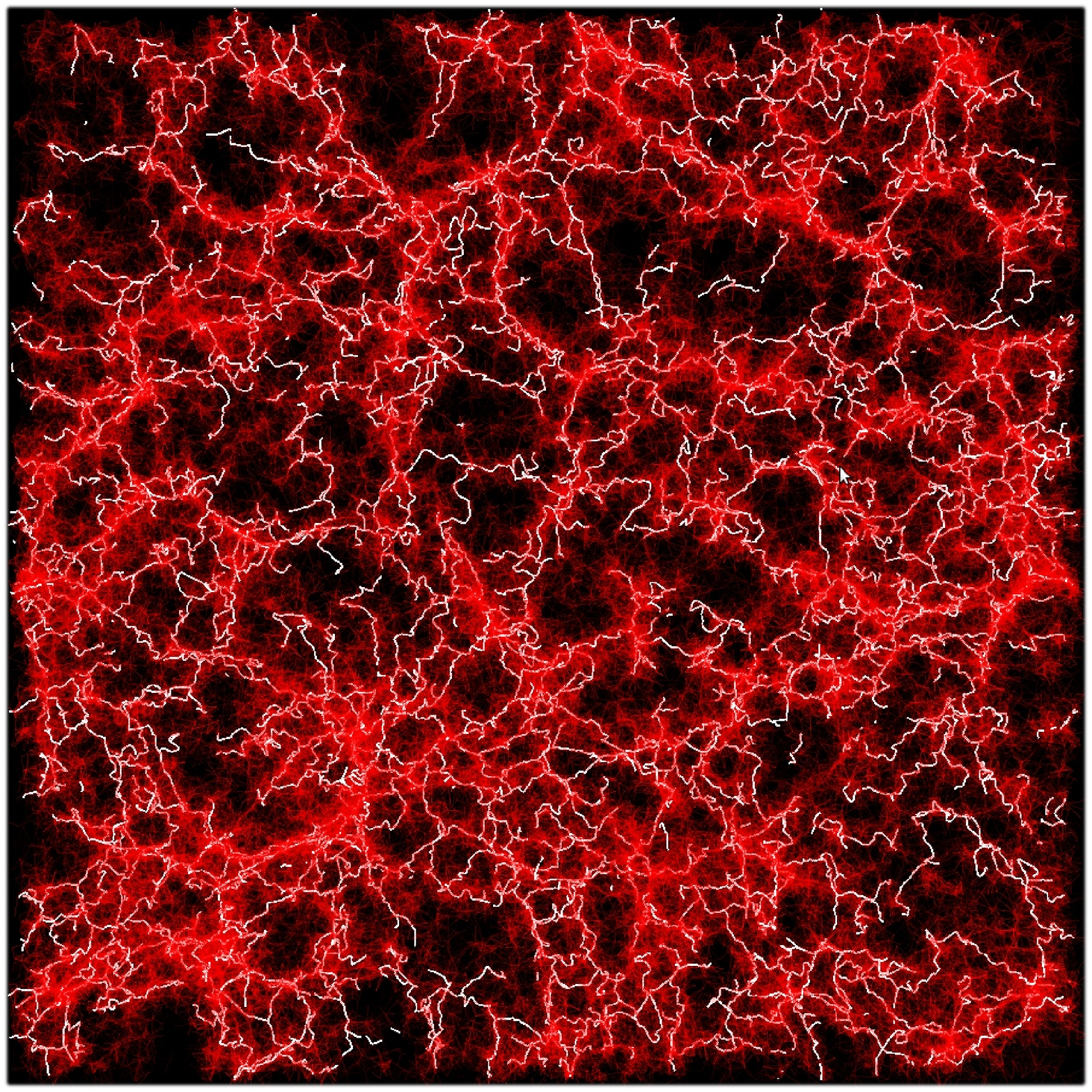}\\
\centering\includegraphics[height=0.28\textheight]{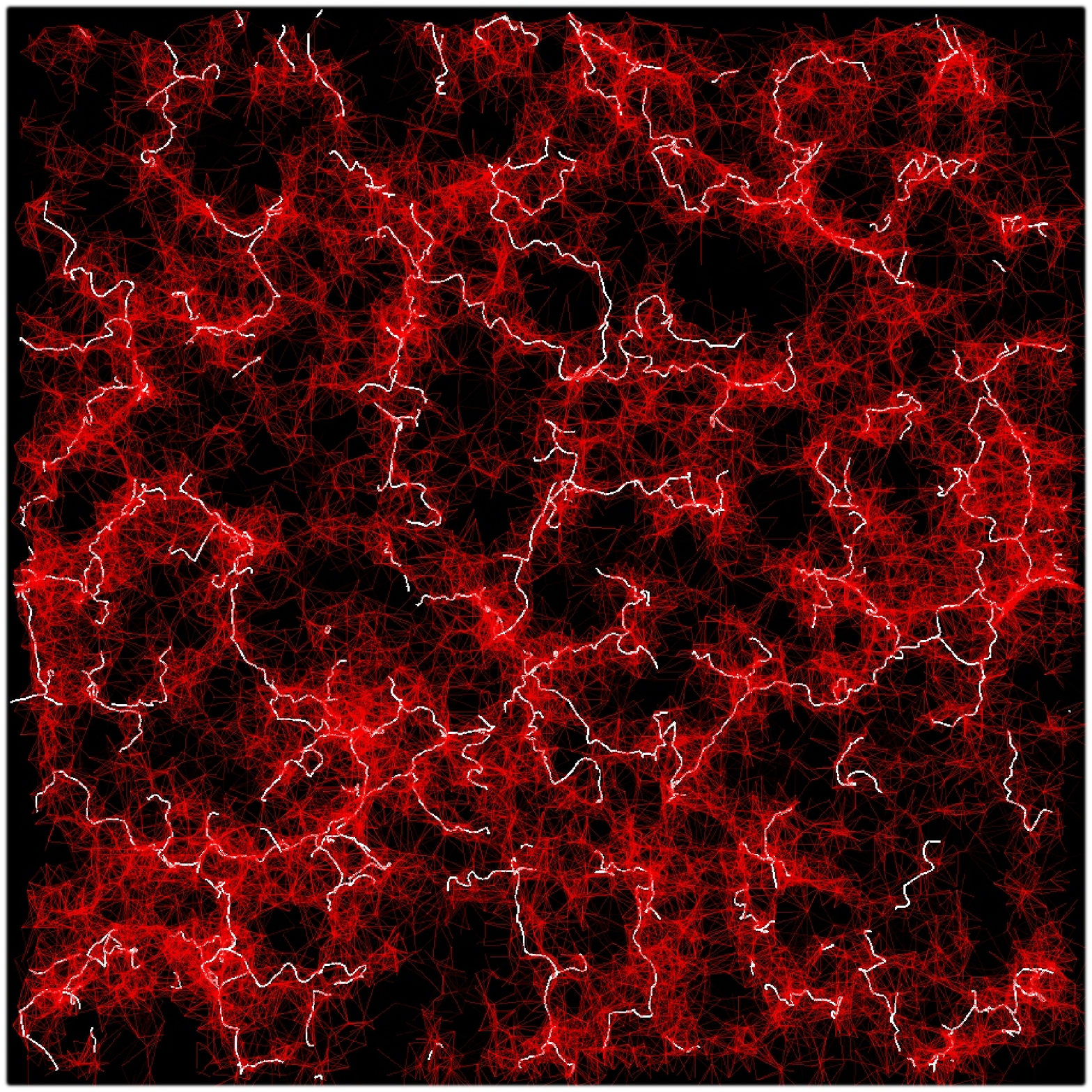}\\
\centering\includegraphics[height=0.28\textheight]{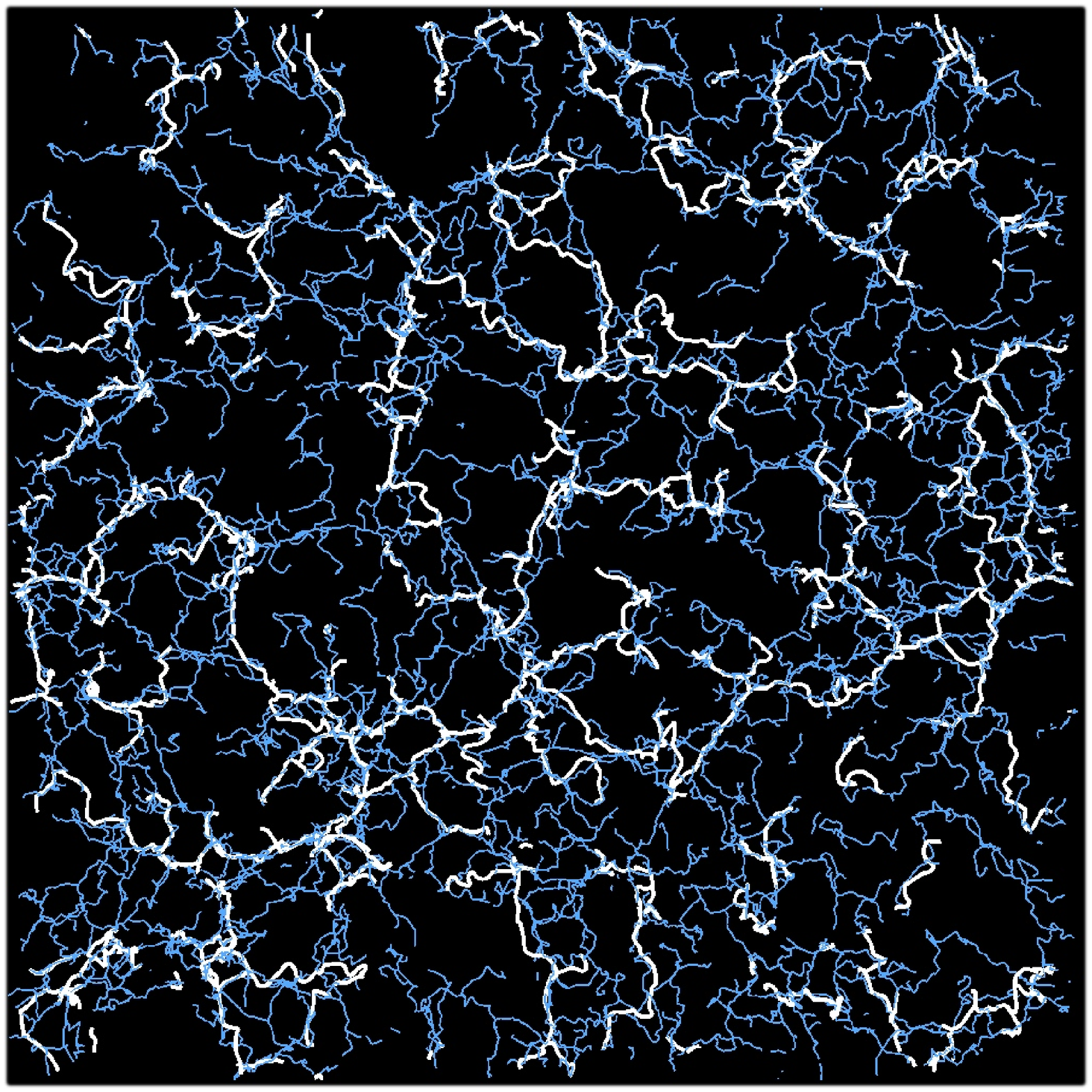}\\
\caption{The filamentary distribution above a \hyperref[defpers]{persistence} level of $\nsig{4}$ in a $250\times250\times20\Mpc$ slice of a $512^3$ particles and $250\Mpccc$ large cosmological simulation. The red segments on the {\sl top} and {\sl central } figures correspond to the segments of the Delaunay tesselation of a $128^3$ and $64^3$ sub-sample, on which the corresponding filaments have been computed. On the {\sl bottom} figure, the thick white filaments correspond to the $64^3$ sub-sample while the blue thin filaments where computed on the $128^3$ sub-sample. This figure clearly demonstrates that \progname\, is able to grasp the main features of the cosmic web with relatively sparse sample.
\label{fig_simuD_resolution} }
\end{minipage}
\end{figure}

Let us finally address a straightforward question: to what extend does \progname\, manage to grasp the main features of the cosmic web with relatively sparse samples? 
Figure~\ref{fig_simuD_resolution} illustrates this query while comparing the filaments computed from two sub samples of variing resolution of from a $250\Mpccc$ large cosmological simulation with   $512^3$ particles (namely  $64^3$ sub-sample and $128^3$ sub-sample respectively). From this figure, it seems that indeed, the features which are identified in the 
sparser sample are real, since they are also found in the more densely sampled catalogue. There seems to be some encouraging level of convergence between the two sets of critical lines.

\subsection{Persistent peak identification}

From visual inspection, it therefore seems relatively clear that the technique developed in this paper is able to correctly decompose the cosmic web into simpler objects of astrophysical interest. However, the approach is based on one fundamental assumption, which is that the ascending and descending \hyperref[defmanifold]{manifolds} of Morse theory, each associated to a specific type of critical point, are representative of the voids, filaments, walls and haloes. While the astrophysical nature of a filament or a wall is not defined very precisely, but is rather understood intuitively, this is not the case of a dark matter halo for instance, which is supposed to be a gravitationally bound structure and the fact that the persistent maxima of the density field correctly identify the gravitationally bound structures is not established. We check this assumption by comparing the distribution of dark matter haloes identified by a simple friend of friend technique (FOF hereafter, see \citet{FOFinfo} for instance) in a $100\Mpccc$, $512^3$ $\lcdm$ dark matter simulation to the \hyperref[defpers]{persistence} diagram in the same simulation, as illustrated on figure \ref{fig_halo_per_diag}.\\

On this figure, the probability distribution function (PDF) of the \hyperref[defpers]{persistence} pairs\footnote{As in section \ref{subsec_sigthres}, each pair is represented by a point with coordinates the density of each of its critical point, see that section for more explanations.} of type $2$ (\ie the maxima/$1$-saddle points pairs) measured in a $128^3$ particles sub-sample is displayed in the density/density plane, the horizontal axis corresponding to the density of the $1$-saddle point, and the vertical one to that of the maximum. The green line therefore represents the $0$-\hyperref[defpers]{persistence} limit, while the oblique white dashed and dotted lines delimits the $\nsig{4}$ and $\nsig{5}$ threshold respectively. In order to compare this distribution to that of the astrophysical dark matter haloes, each of them is also represented as a disk with coordinates that of the \hyperref[defpers]{persistence} pair of its most dense particle (the densest particle within a halo is necessarily a local maximum). Each halo was identified using a standard linking length parameter of one fifth of the average inter particular distance, and the red disks represent the haloes with mass $M>73.8\times10^{10}\;M_\odot$ (\ie with more than $1,280$ particles in the initial simulation, or $20$ in a $128^3$ sub sample), while the green ones represent the haloes with mass $M>590.4\times 10^{10}\;M_\odot$ (\ie with more than $10,240$ particles in the initial simulation, or $1,280$ in a $128^3$ sub sample). It is a very striking result how well the population of dark matter halos is localized in the \hyperref[defpers]{persistence} diagram. While lighter ones (red disks) mostly correspond to maxima with \hyperref[defpers]{persistence} ratio higher than $\nsig{4}$ and overdensity $\rho/\rho_0 > 4\times10^3$, the heavier ones lie in the zone with \hyperref[defpers]{persistence} higher than $\nsig{5}$ and overdensity $\rho/\rho_0 > 3.2\times10^4$.\\ 

These results mean that \hyperref[defpers]{persistence} selection associated to a global overdensity threshold is naturally (\ie without any specific qualibration) a very good halo finder,  which is quite encouraging, and validates  our initial assumption on the relationship between the persistent topological features and the astrophysical components of the cosmic web. This is illustrated on figure \ref{fig_simu_halos_pic} where each dark matter halo with mass $M>73.8\times10^{10}\;M_\odot$ (\ie the red disks of figure \ref{fig_halo_per_diag}) is colored in blue. Once again, it is clear on the central frame that all haloes along large filaments are correctly linked by the \hyperref[defDMC]{DMC}. We also remark that the \hyperref[defDMC]{DMC} and \hyperref[defpers]{persistence} pairs contain unexploited information of the topology as our algorithm explicitly identify the $k$-cycles as sequences of critical points associated to \hyperref[defpers]{persistence} pairs (see \paperone). For instance, each \hyperref[defpers]{persistence} pair associated to a halo correspond to a $0$-cycle that define a principal filament, as shown on the bottom frame, where only the filaments corresponding to \hyperref[defpers]{persistence} pairs whose maximum is a dark matter halo are represented. Moreover, using the information contained in the persistence pairs, one basically obtains a hierarchical structure finder that is able to also identify substructures not only within clusters, but also within filaments, walls and voids.    

\section{Our universe: the SDSS catalogue}
\label{sec_SDSS}
Let us now  illustrate a few prospective measurements of the filamentary structure of the actual galaxy distribution in the Universe. The ultimate goal of such measurements is to allow a complete and precise characterization of the properties of the filamentary structure of the galaxy distribution by measuring their topological properties, such as the Betti number and Euler characteristics, and modeling the geometrical characteristics of the voids, walls and filaments (\ie their total length, number, the number of filaments per galaxy clusters, ...). 
Such a task is rather challenging, as it requires the construction of realistic mock observations from N-body simulations to asses the influence of observational biases and distortions; it also requires a lot of care in the handling of the observational data themselves (for instance by taking into account the complex survey geometry, among other difficulties).  In this paper, we focus on  convincing the reader that the method we introduced  paper I is particularly suited to such a task by showing how easily and efficiently it can be applied to a real galaxy catalogue. We postpone the full investigation to a future paper.

\subsection{The cosmic web in the SDSS}

\begin{figure}
\centering
\includegraphics[width=\linewidth]{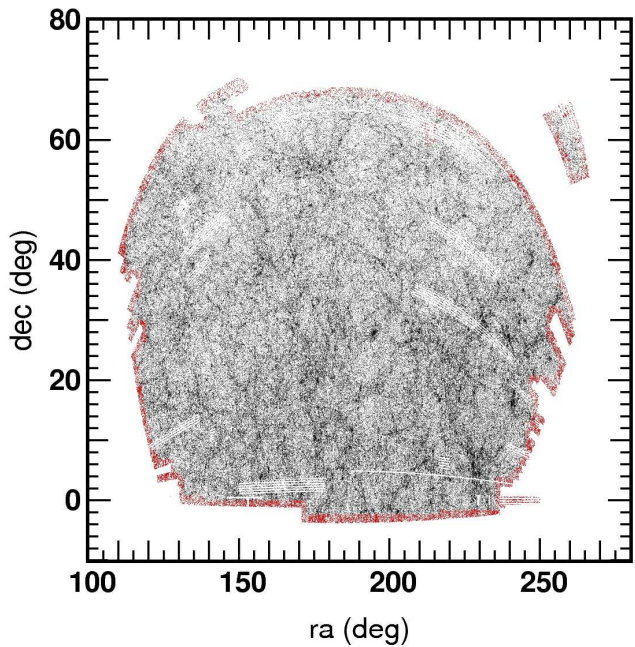}
\caption{Angular distribution of the $515,458$ galaxies corresponding to a sub-sample of the SDSS DR7 galaxy catalogue that we use in our tests (see main text for selection criterion). The $66,608$ red galaxies are those detected as being on the boundary of the distribution using the method described in the main text. Note that some regions were not fully scanned and exhibit series of thin empty parallel stripes, but we simply ignore that fact when computing the boundaries.\label{fig_SDSS_radec}}
\end{figure}

For that purpose, we use data from the $7^{\rm th}$ data release (DR7) of the SDSS \citep{Abazajian09}, and in particular the large-scale structure subsample called {\em dr72bright0} sample of the New York University Value Added catalogue \citep{blanton05}, which is made of a spectroscopic sample of galaxies with u,g,r,i,z- band (K-corrected) absolute magnitudes, r-band apparent magnitude $m_r$, redshifts, and information on the mask of the survey. In that sample, the spectroscopic galaxies are originially selected under the conditions that $10.0 \leq m_r \leq 17.6$ and $0.001 \leq z \leq 0.5$, but we further cut the sample for the purpose of our tests, restraining it to the galaxies with $z \leq 0.26$ and right ascension $100\deg\leq RA \leq 280\deg$, which removes the three thin stripes in the southern Galactic hemisphere. The resulting angular distribution, containing $515,458$ galaxies among the $567,759$ in the original sample is displayed on figure \ref{fig_SDSS_radec}.\\

\begin{figure}
\centering
\includegraphics[width=\linewidth]{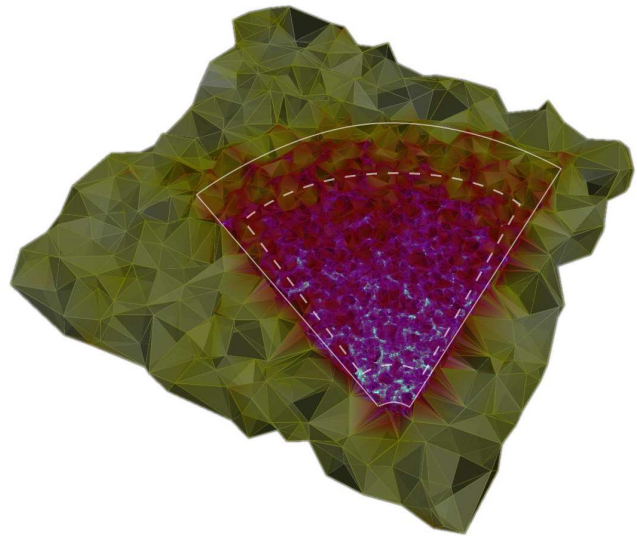}
\caption{A slice within the delaunay tesselation of the distribution used to compute the \hyperref[defDMC]{DMC} of the SDSS. The plain white contour delimits the SDSS distribution (inside) and the randomly added low density particles that fill the void regions of the bounding box (outside). Any galaxy outside the white dashed contours is considered as being on the boundary.\label{fig_SDSS_select}}
\end{figure}

In order to compute the \hyperref[defDMC]{DMC} of the observed galaxy distribution, we will use the mirror type boundary conditions as introduced in paper I. This type of boundary conditions normally apply to distributions enclosed within parallelepiped boxes, which is not the case here. In the simple case of a box-like geometry, the particles within a given distance of the faces are mirrored, and any particle outside the initial box or whose DTFE density may be affected by the content of the exterior of the box is tagged as a boundary particle. As the geometry of the SDSS catalogue is complex, we simply enclose it within a slightly larger box, fill the empty regions with a low density random distribution of particles, and then mirror the boundaries. The mirrored particles and the random ones are tagged and we then identify the boundaries of the galaxies distribution and tag as well those galaxies whose DTFE density may depend on the distribution outside the observational region. Although the catalogue does contain precise information about the mask of the survey, we prefer to use a simple though  automatic method to identify the boundaries of the galaxy distribution. This method simply  samples the angular galaxy distribution in the RA/DEC plane over a regular grid of $1\times 1\deg$ pixels, and identifies the galaxies on the boundary of the catalogue as those that belong to a pixel with at least one completely empty neighbor. Note that such a method presents the advantage of being generic, as it does not presume any previous knowledge of the mask, and could therefore be applied directly to other galaxy catalogues. The resulting boundary galaxies are represented in red on figure \ref{fig_SDSS_radec}. We finally also tag those galaxies with redshift $z \leq 0.02$ and $z \geq 0.2$ as boundary and proceed with the computation of the \hyperref[defDMC]{DMC}, as in the regular mirror type boundary condition case. A slice of the Delaunay tesselation of the final distribution is displayed on figure \ref{fig_SDSS_select}.\\

\begin{figure}
\begin{minipage}[c][\textheight]{\linewidth}
\includegraphics[width=\linewidth]{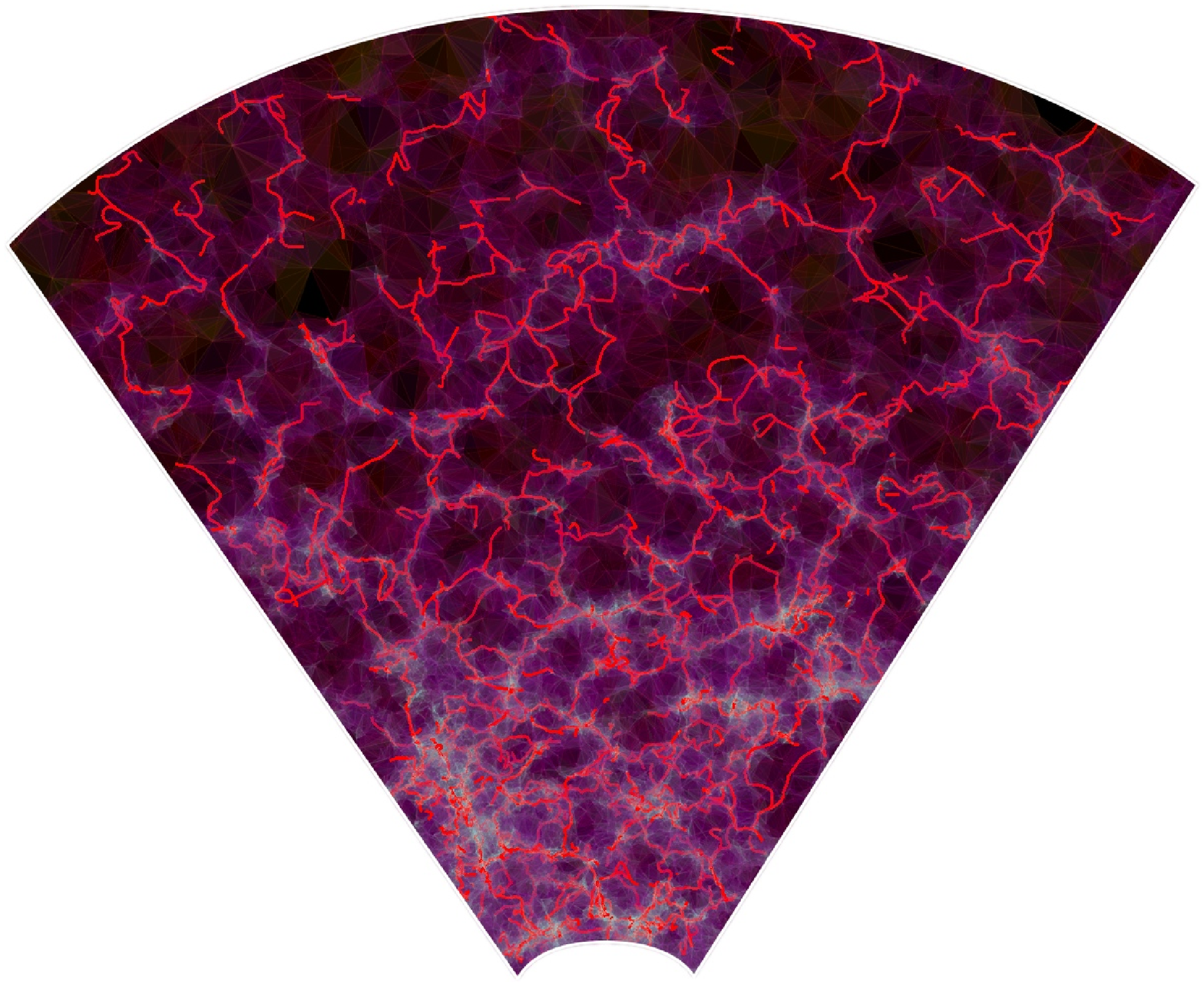}\\
\includegraphics[width=\linewidth]{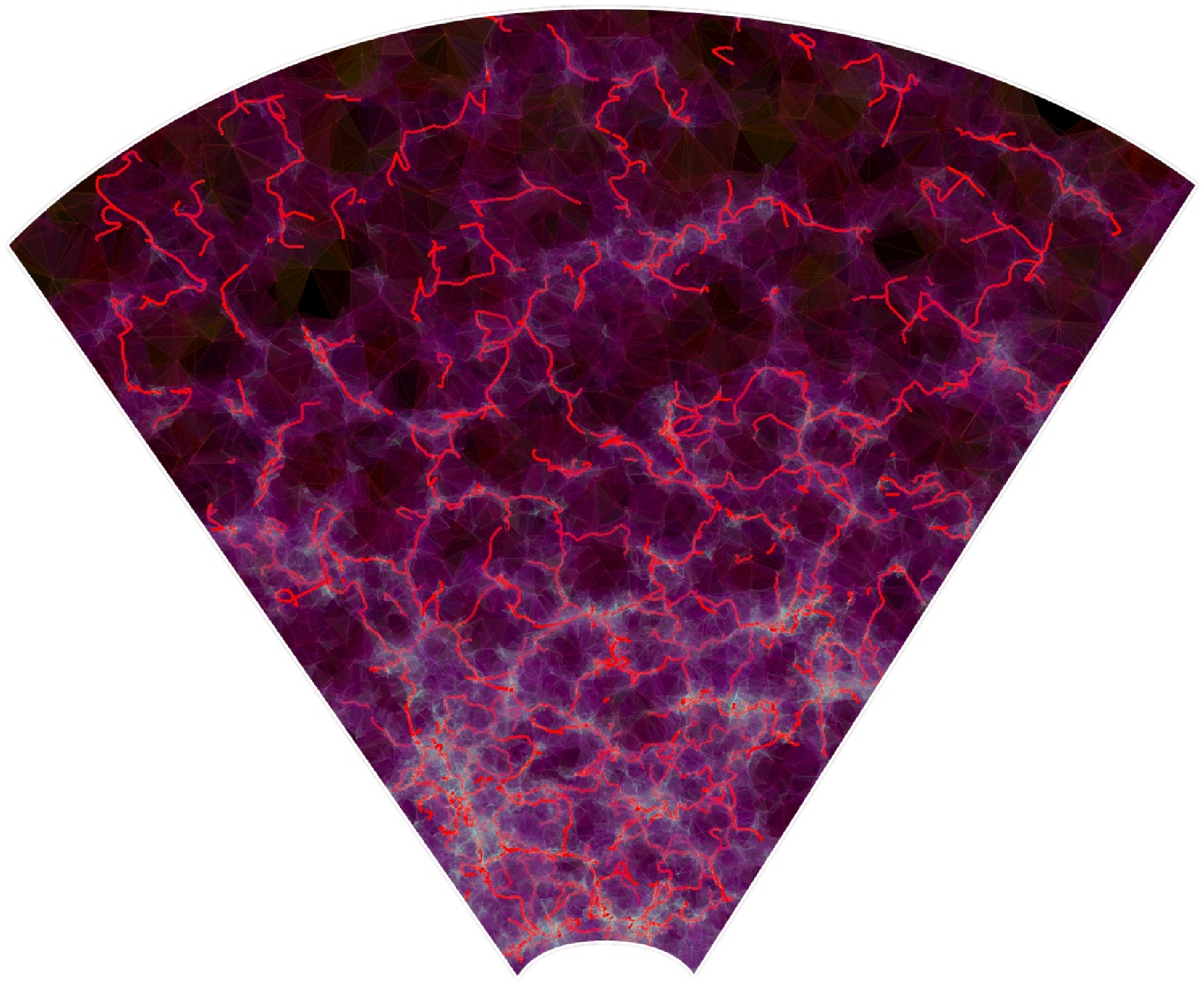}\\
\includegraphics[width=\linewidth]{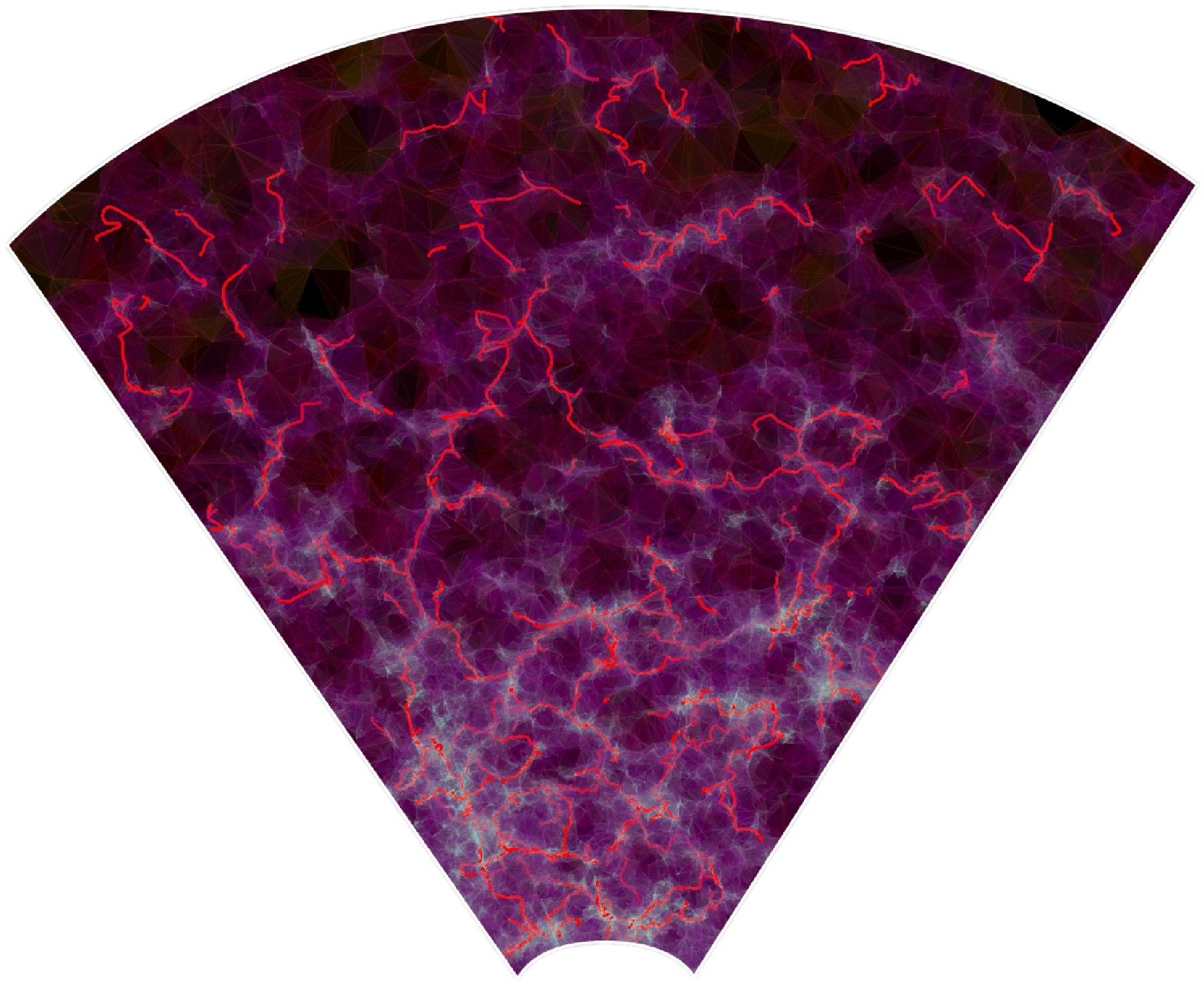}\\
\caption{ From top to bottom, the filamentary structure in a $\sim40\Mpc$ thick slice of the SDSS Dr7 galaxy catalogue at a significance level of $3$, $4$ and $\nsig{5}$ respectively. The distribution is represented by the non bounding subset (see main text) of the Delaunay tesselation used to compute the \hyperref[defDMC]{DMC}, shaded according to the logarithm of the density. The depth of a filament can be judged by how dimmed its shade is. Note that filaments that seem to stop for no apparent reason actually enter or leave the slice.\label{fig_delaunaySDSS}}
\end{minipage}
\end{figure}

\begin{figure*}
\begin{minipage}[c][\textheight]{\linewidth}
\centering\subfigure[A portion of SDSS DR7]{\includegraphics[width=0.32\linewidth]{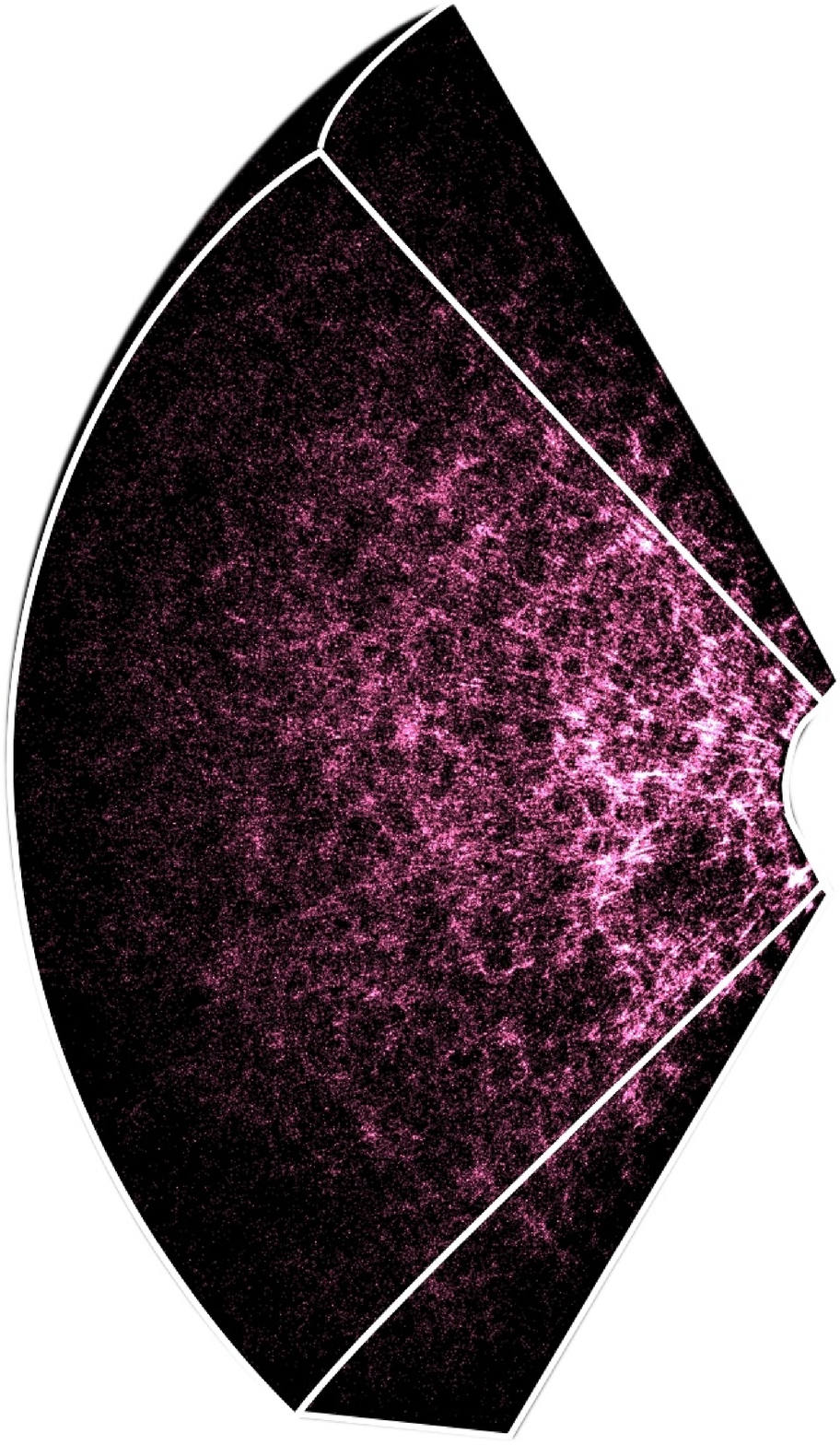}\label{fig_SDSS_picA}}
\hfill  \subfigure[The filamentary structure]{\includegraphics[width=0.32\linewidth]{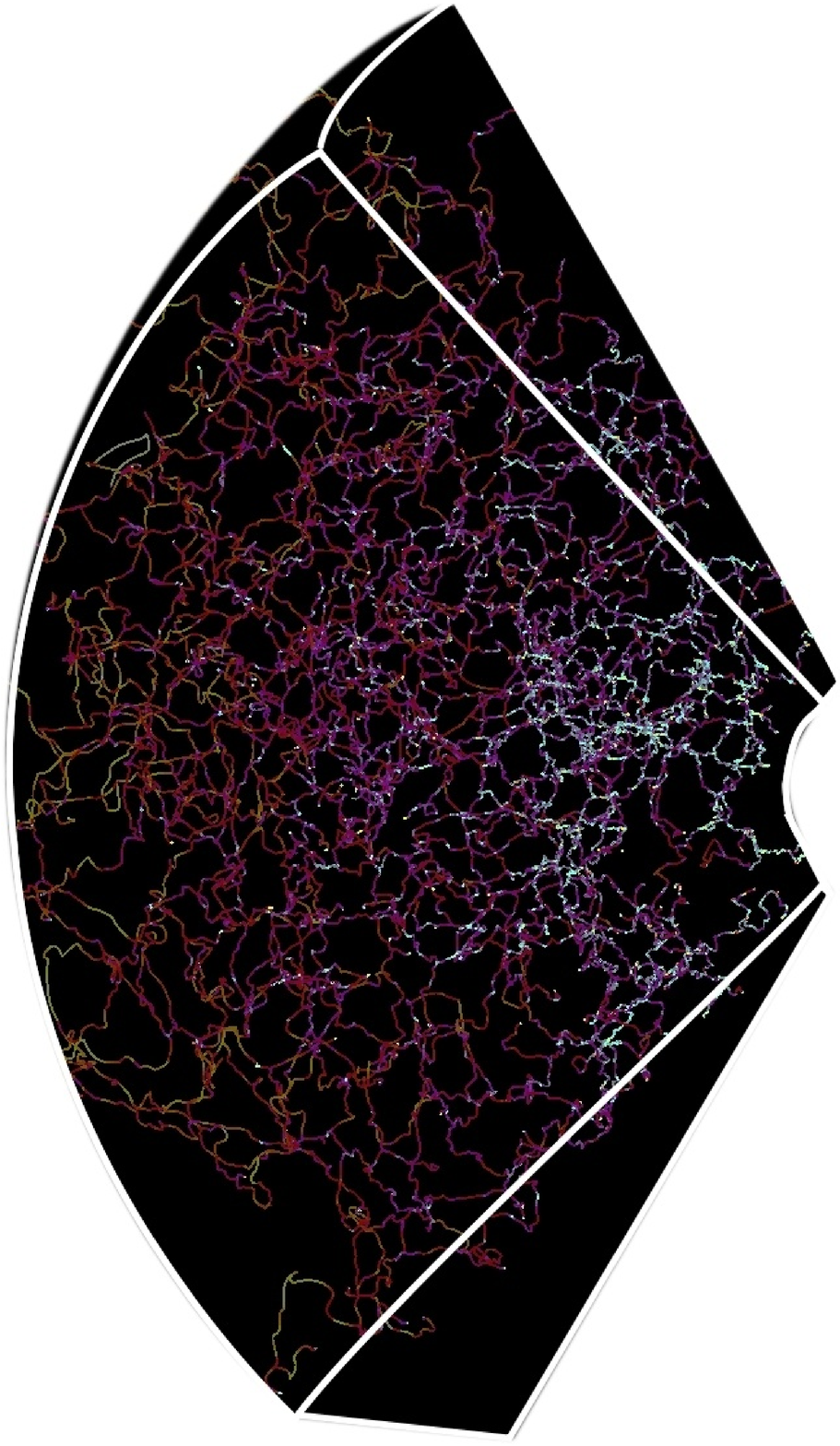}\label{fig_SDSS_picB}}
\hfill\subfigure[Three voids]{\includegraphics[width=0.32\linewidth]{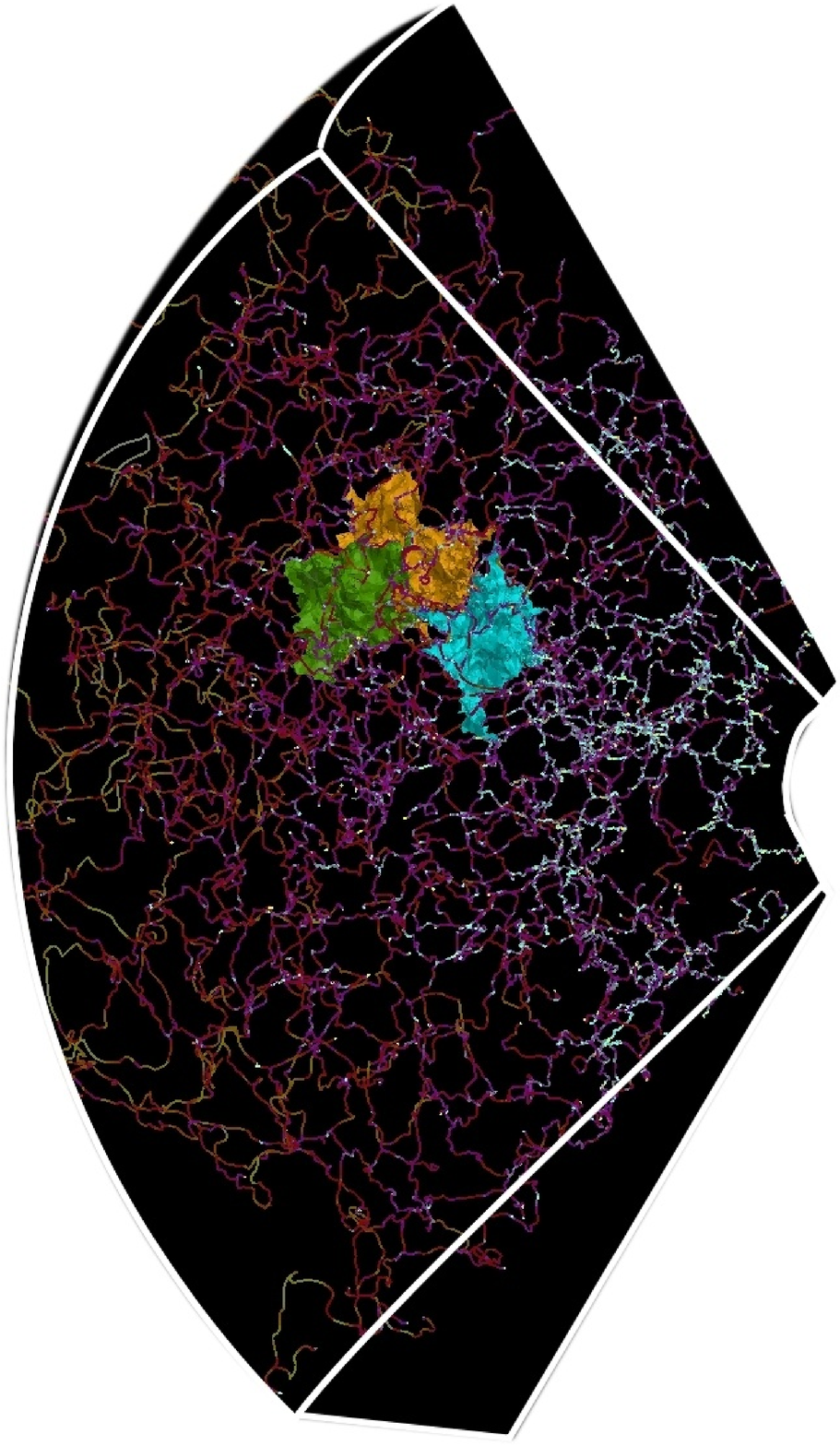}\label{fig_SDSS_picC}}\\
\centering  \subfigure[A zoom on the voids and the filamentary structure]{\includegraphics[height=0.4\textheight]{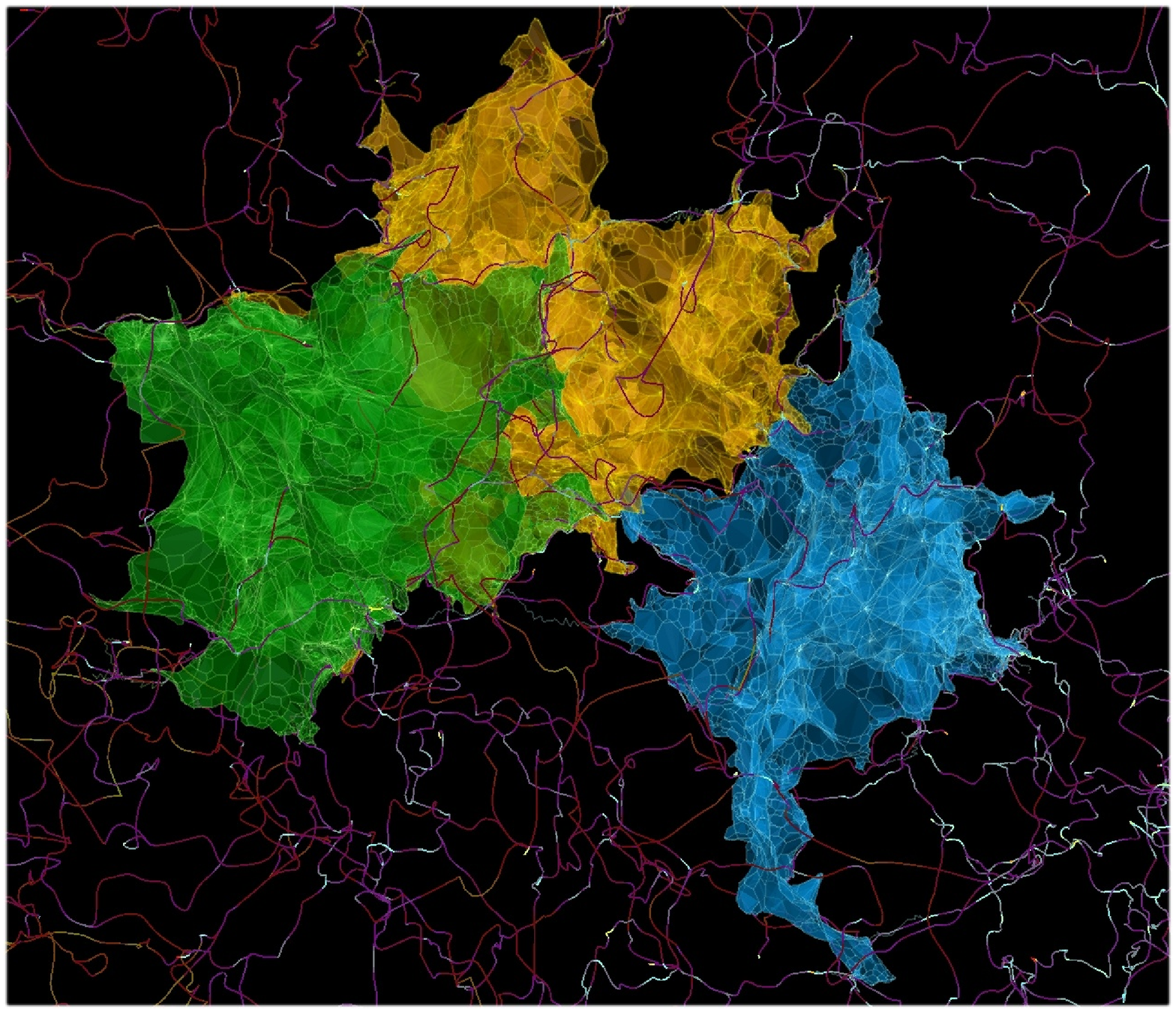}\label{fig_SDSS_picD}}\\
\caption{ The detected filamentary structure at a significance level of $\nsig{5}$ and three voids within a portion of SDSS DR7. Note that only the upper half of the distribution shown on figure \ref{fig_SDSS_radec} is displayed here for clarity reasons. The color of the filaments corresponds the the logarithm of the density field. The filaments of the SDSS extracted with \progname\, is readily available online at the URL {\em\tt http://www.iap.fr/users/sousbie/}.\label{fig_SDSS_pic} }
\end{minipage}
\end{figure*}

\begin{figure*}
\begin{minipage}[c][\textheight]{\linewidth}
\begin{minipage}[c]{0.49\linewidth}
\centering\includegraphics[height=0.3\textheight]{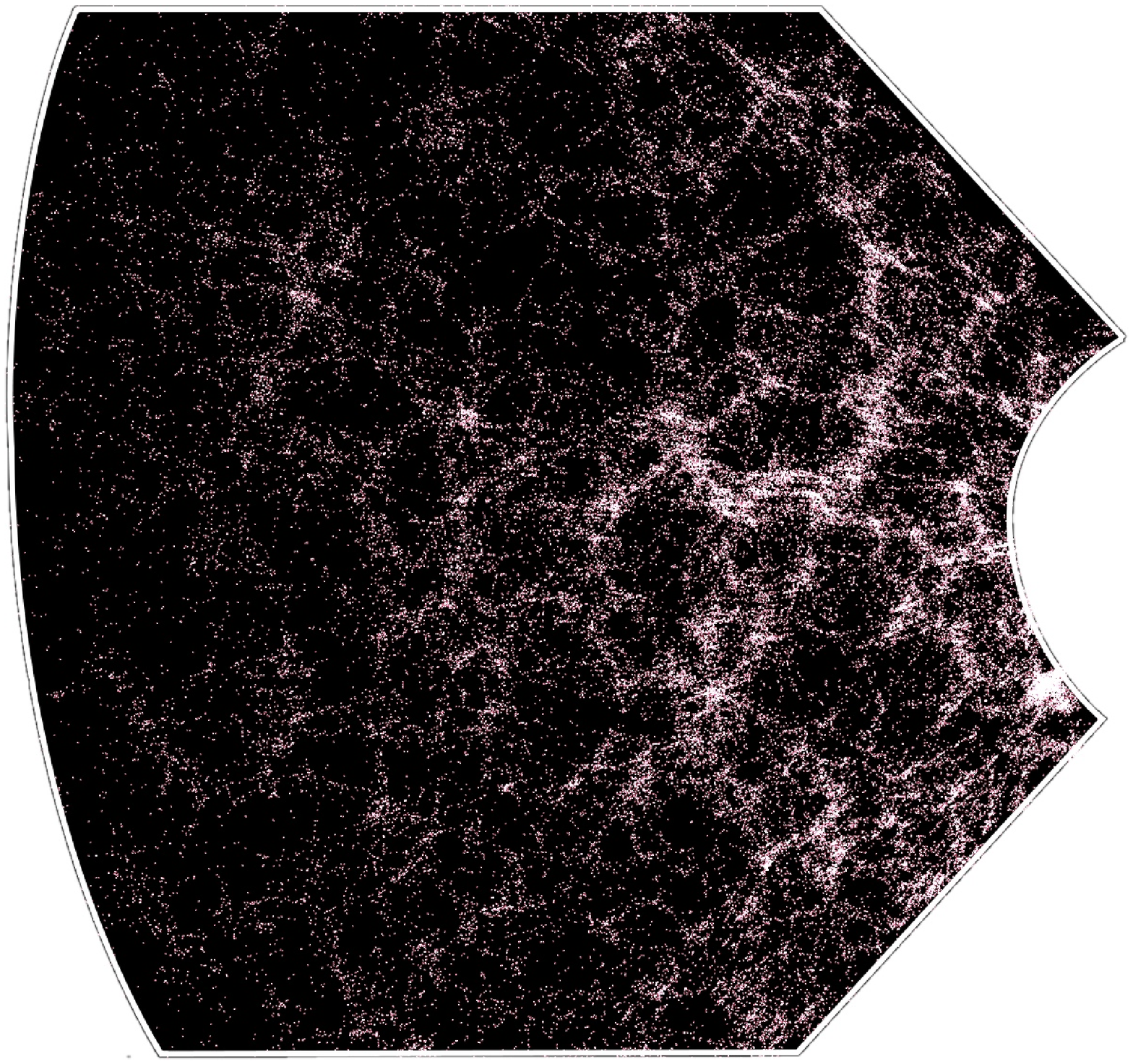}\\
\centering\includegraphics[height=0.3\textheight]{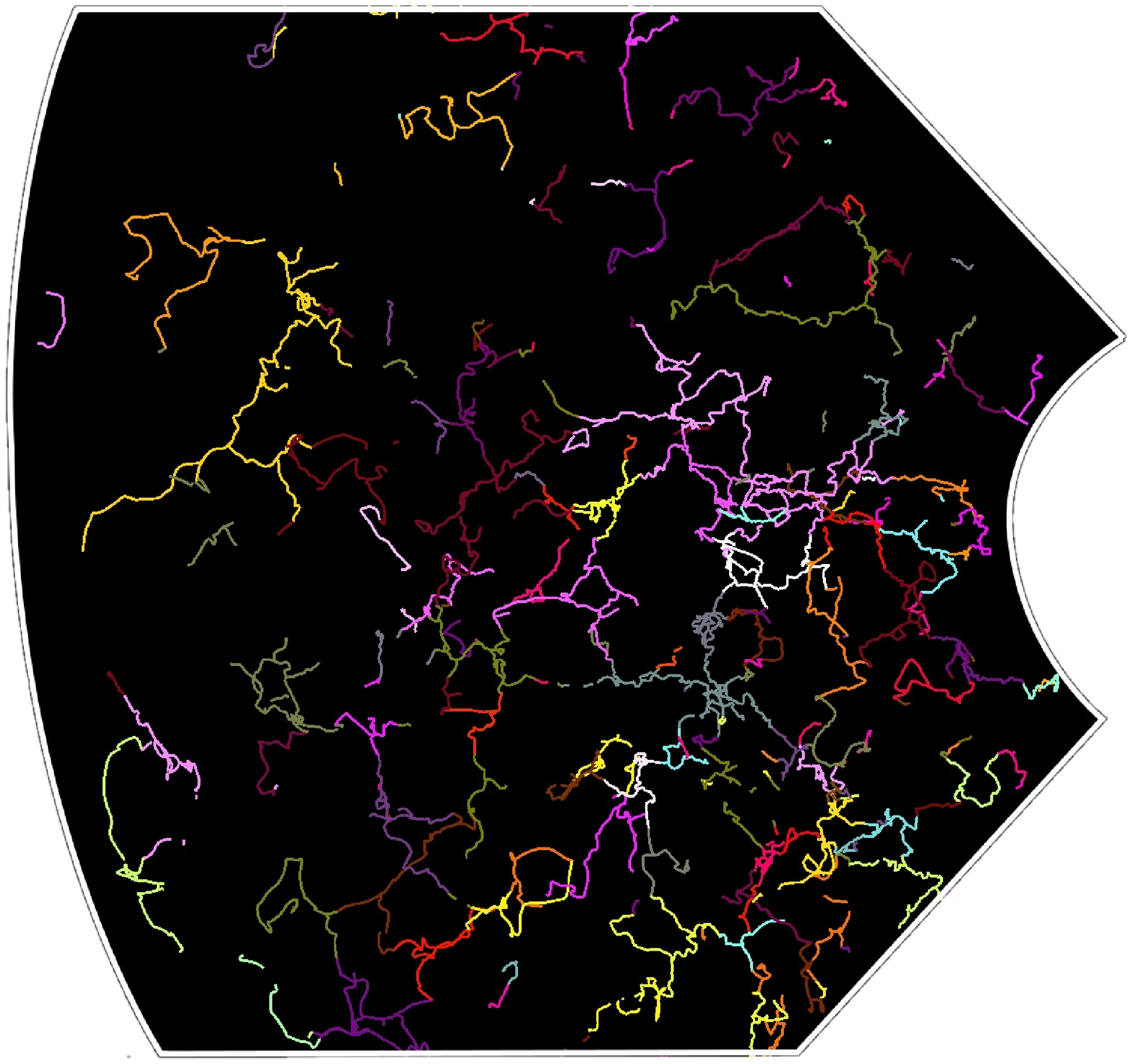}\\
\centering\includegraphics[height=0.3\textheight]{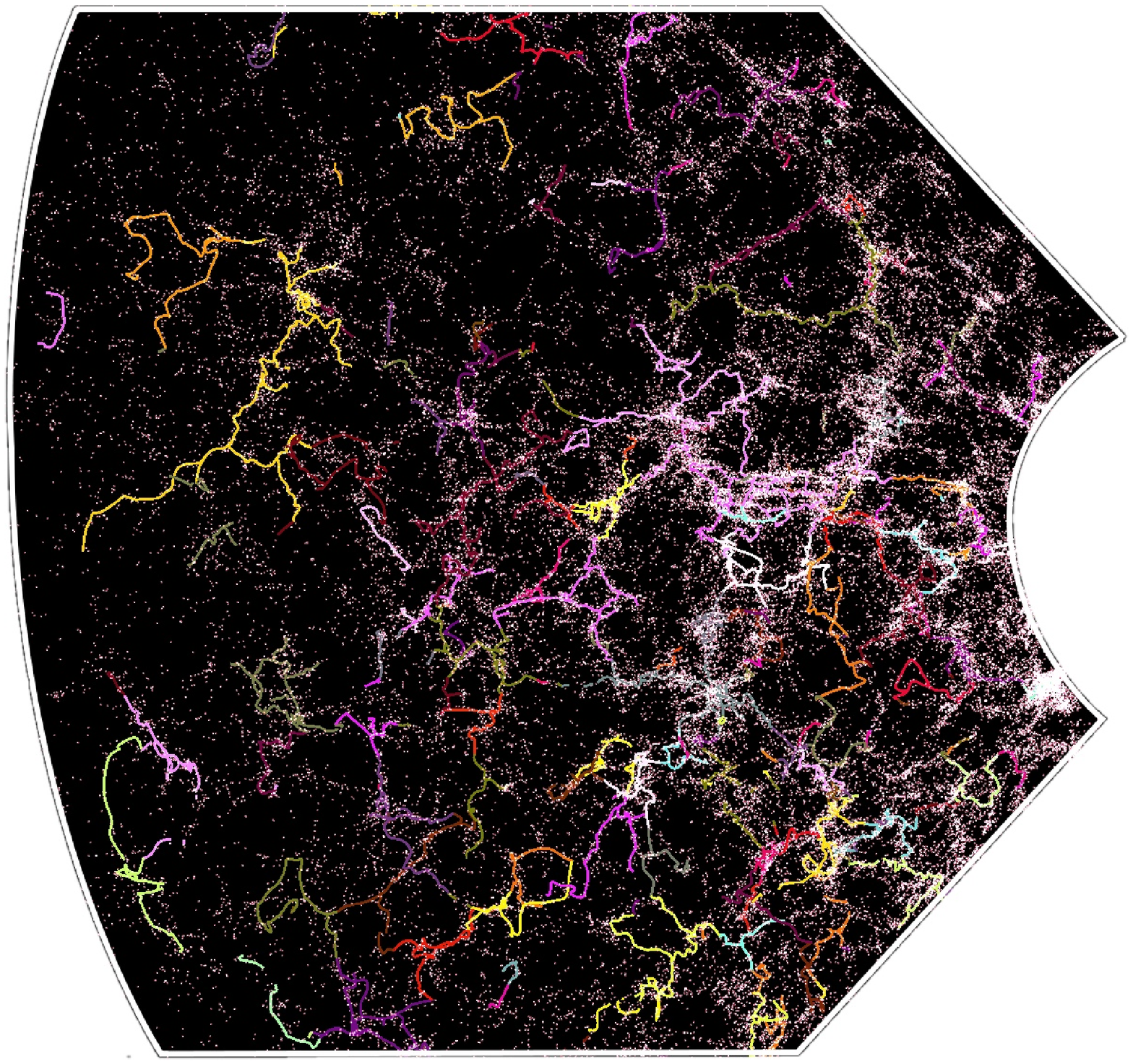}\\
\end{minipage}
\hfill
\begin{minipage}[c]{0.49\linewidth}
\centering\includegraphics[height=0.3\textheight]{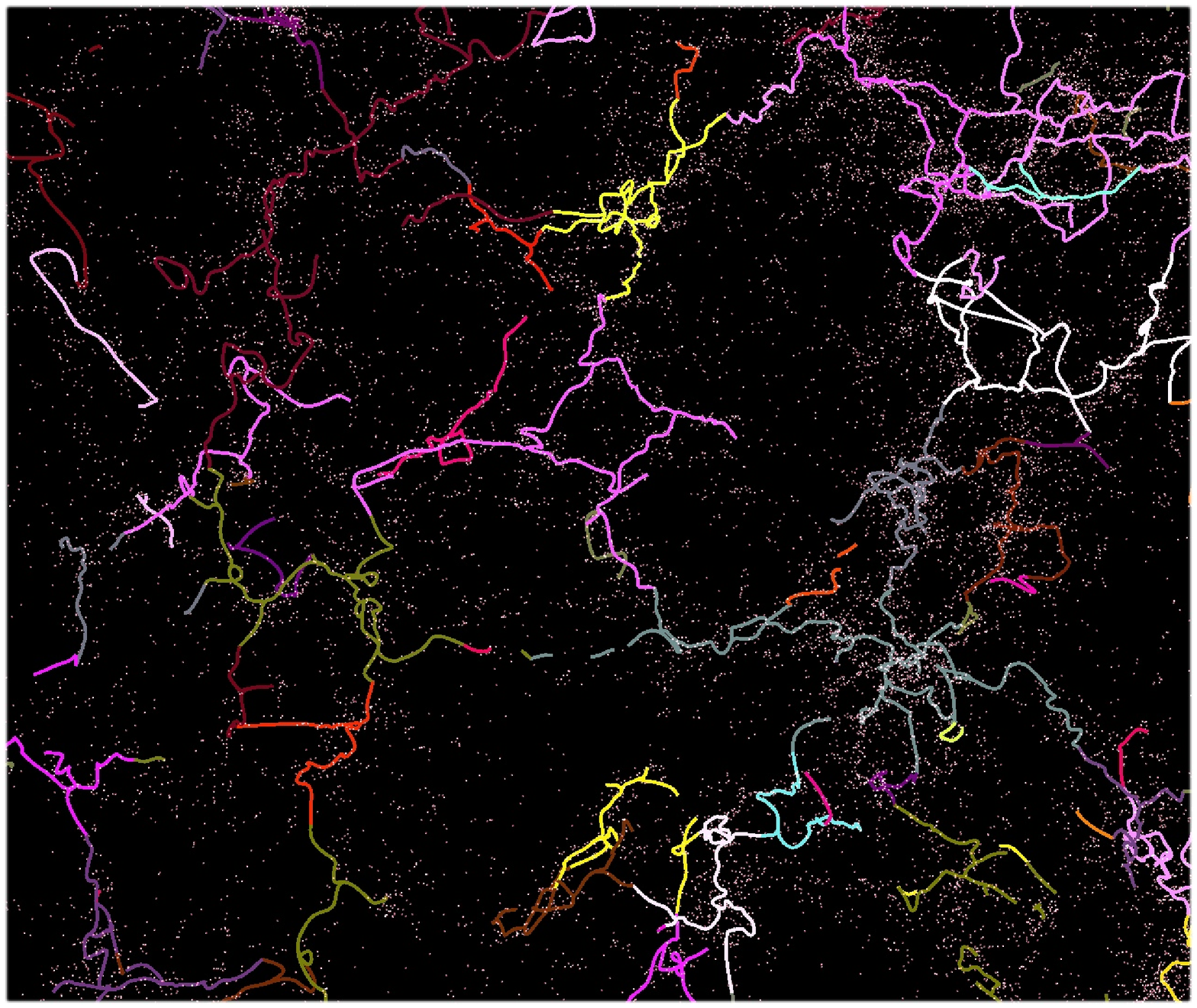}\\
\centering\includegraphics[height=0.3\textheight]{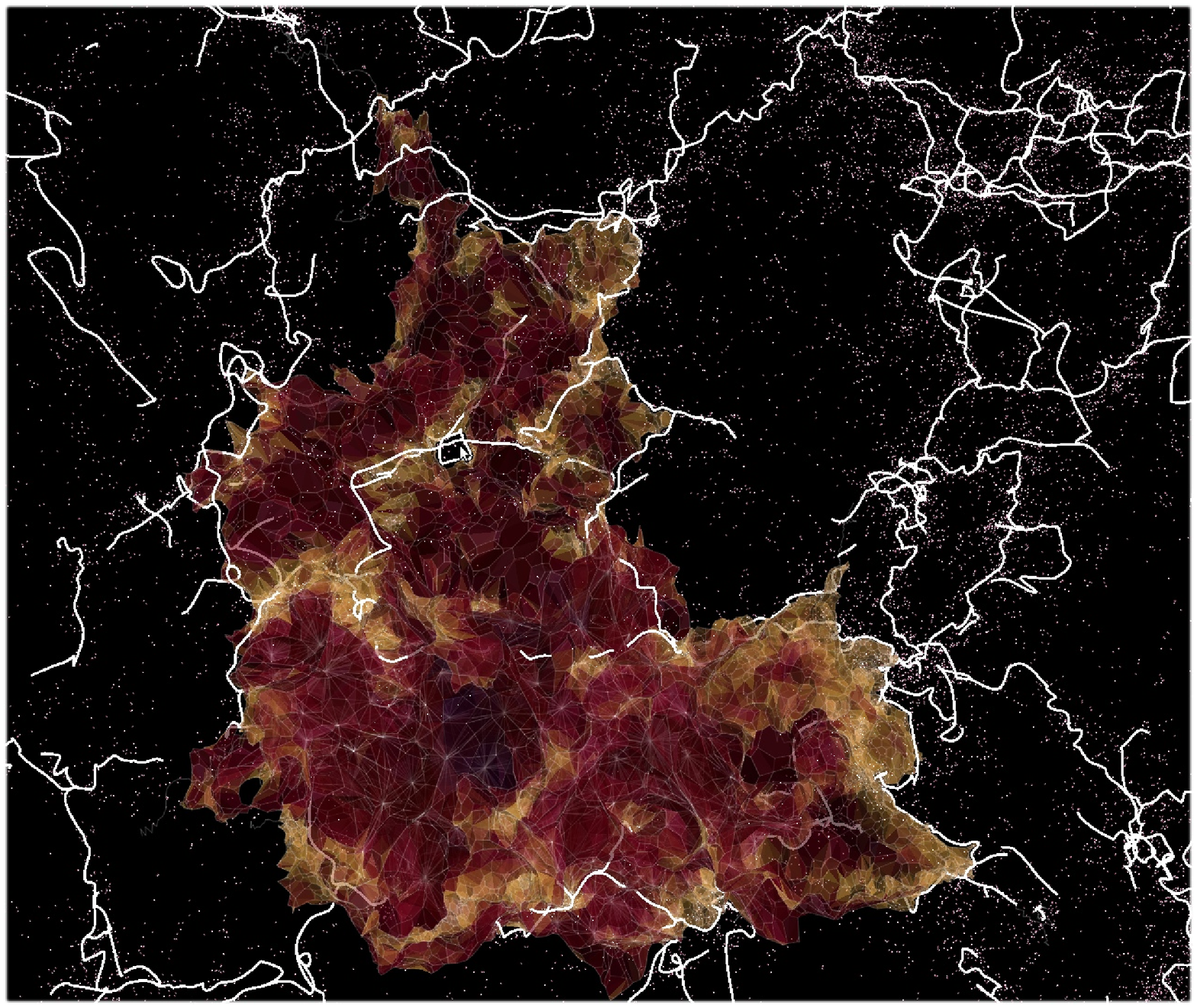}\\
\centering\includegraphics[height=0.3\textheight]{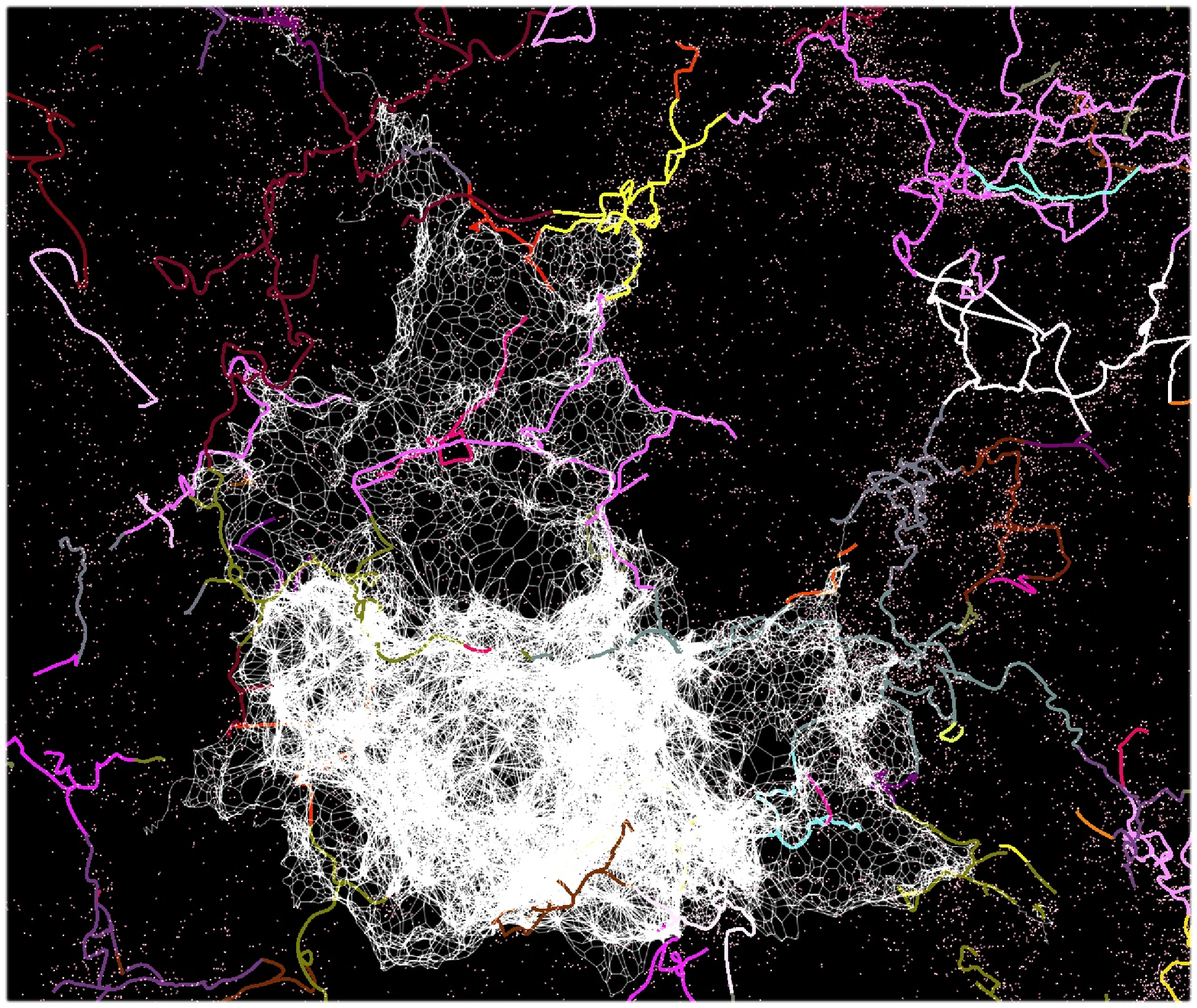}\\
\end{minipage}
\caption{The filamentary structure (left) and a void (right) detected at a significance level of $\nsig{5}$ in SDSS DR7. In order to emphasize the filamentary structure, only a $\sim 60\Mpc$ thick flat slice of the distribution is displayed on each frame. The void surface is shaded according to the log of the density field (central right frame), while the color of each \hyperref[defarc]{arc} of the \hyperref[defDMC]{DMC} tracing the filamentary structure depends on the index of the maximum to which it is connected. Note that the foremost part of the voids on the central and bottom right picture protrudes from the slice, while the filaments are trimmed to its surface.
Given its shape, this  void is in fact a good  example of why we should identify filaments via a DMC rather than using a Watershed technique, as it displays two strong ``thin wings"
which would lead to the incorrect detection of spurious sets of boundaries.
\label{fig_SDSS_slice}}
\end{minipage}
\end{figure*}

The resulting \hyperref[defDMC]{DMC} covers the $440,950$ galaxies in black on figure \ref{fig_SDSS_radec} and obeying the additional condition $0.02\leq z \leq 0.2$ (or equivalently $85\leq d \leq 860\Mpc$) and it is displayed on figures \ref{fig_delaunaySDSS}, \ref{fig_SDSS_pic}, and \ref{fig_SDSS_slice}. Figure \ref{fig_delaunaySDSS} illustrates the influence of the significance level on the measured filamentary network. On this figure, the filaments (\ie the ascending  $1$-\hyperref[defmanifold]{manifolds} or \hyperref[defarc]{arcs}) within a $\sim 40\Mpc$ slice of the Delaunay tessellations are shown at significance levels of $\nsig{3}$, $\nsig{4}$ and $\nsig{5}$ (from top to bottom); it is quite striking how well more or less significant filaments are accurately identified depending on the value of the \hyperref[defpers]{persistence} ratio threshold. Note how already at a level of $\nsig{3}$ the influence of sampling noise has disappeared and increasing this threshold results in the selection of apparently denser, bigger and longer filaments. As the distant faint galaxies and the nearby bright ones cannot be observed easily, the selection function strongly depends on the distance, and so does the sampling. It reflects in the shade of the Delaunay tessellation, which depends on the logarithm of the density. From a theoretical point of view, the fact that the absolute value of the density is multiplied by the selection function should not affect the detection of the filaments as long as the value of the selection function does not vary much over the typical scale of a filament (or in other word, as long as the topology of the distribution remains unchanged). The measured \hyperref[defpers]{persistence} ratio of \hyperref[defpers]{persistence} pairs may be slightly affected though, when the two critical points in the pair are located at different distances, but this does not seem to have much importance in the present case. A more significant effect results from the scale adaptive nature of DTFE. Because the quality of the sampling decreases with distance, comparatively larger scale filaments are identified as the distance increases and to be able to identify comparable filaments independently of the distance from the observer, one would therefore probably have to resort on volume limited samples.\\

The filamentary structure at $\nsig{5}$ significance level is also shown over larger scales on figure \ref{fig_SDSS_pic} and within a $60\Mpc$ slice where each galaxy is represented by a point on figure \ref{fig_SDSS_slice}. Three voids (\ie ascending $3$-\hyperref[defmanifold]{manifolds}) have been randomly selected within the distribution of figure \ref{fig_SDSS_pic} and are displayed on the bottom frame \ref{fig_SDSS_picD}, showing the intricate relationship between the voids and the filamentary structure that crawls at their surface. As previously observed in simulations, it can be seen on the central right frame of figure \ref{fig_SDSS_slice} that those 3D filaments also trace the 2D filamentary structure at the surface of the voids as expected from Morse theory. Note that it is only because they have been smoothed over four segments to look more appealing and to avoid rendering problems that the filaments do not lie precisely on the surface of the voids. It is in fact a build-in feature of the \hyperref[defDMC]{DMC} and in particular of our implementation that all the different types of identified cosmological structures do form a coherent picture, whatever the properties of the initial discrete sample. This allows for interesting features, such as making possible the count of the number of filaments that belong to a common maxima by intersecting the ascending $1$-\hyperref[defmanifold]{manifolds} with the descending $3$-\hyperref[defmanifold]{manifolds}. This is shown on figure \ref{fig_SDSS_slice} where the color of the filamentary structure corresponds to the index of the maximum it belongs to and individual filaments could be identified the same way, as the two \hyperref[defarc]{arcs} of the \hyperref[defDMC]{DMC} originating from a given saddle point.  

\begin{figure*}
\begin{minipage}[c][]{0.49\linewidth}
\centering\includegraphics[width=\linewidth]{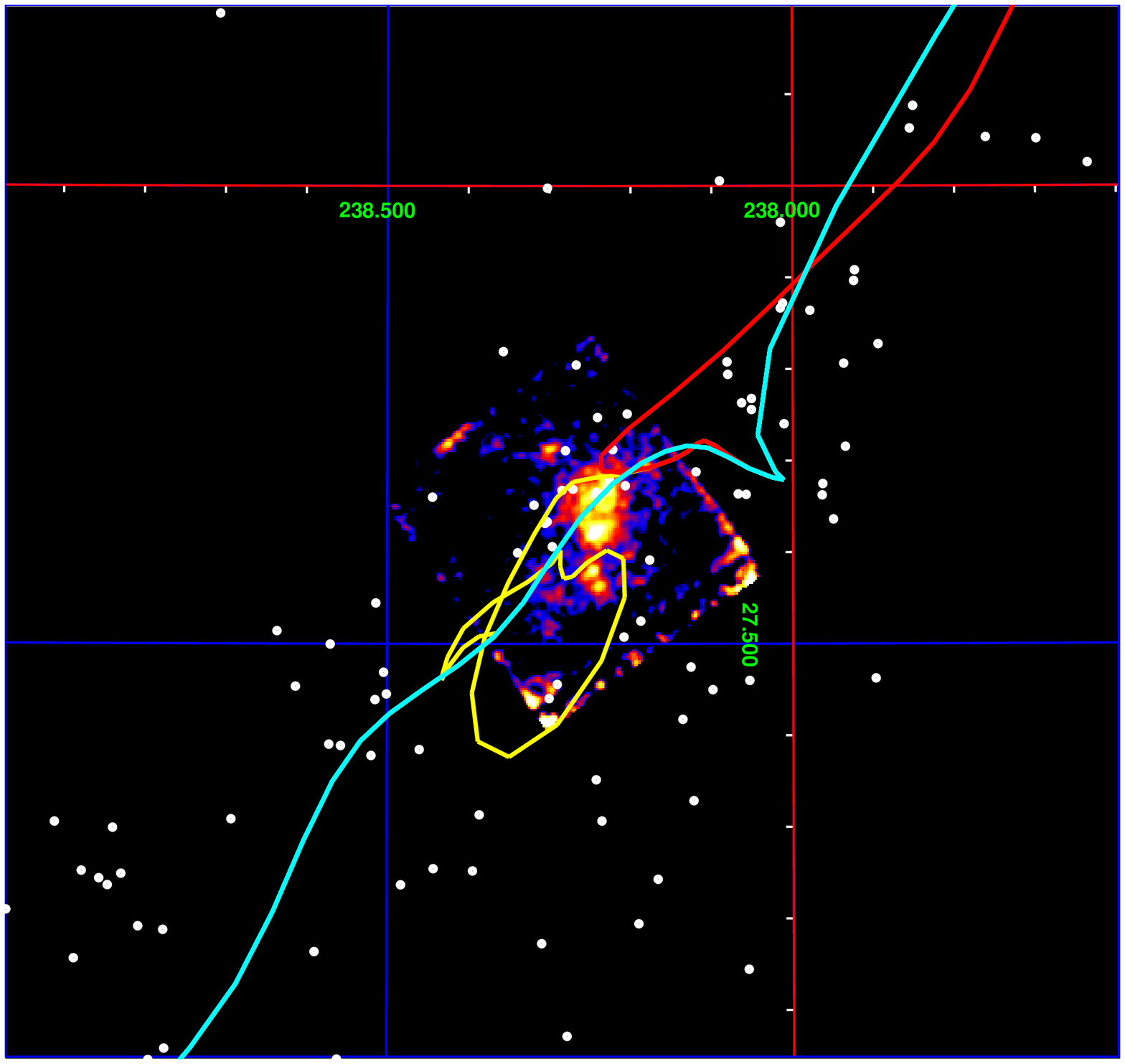}
\end{minipage}
\begin{minipage}[c][]{0.49\linewidth}
\centering\includegraphics[width=\linewidth]{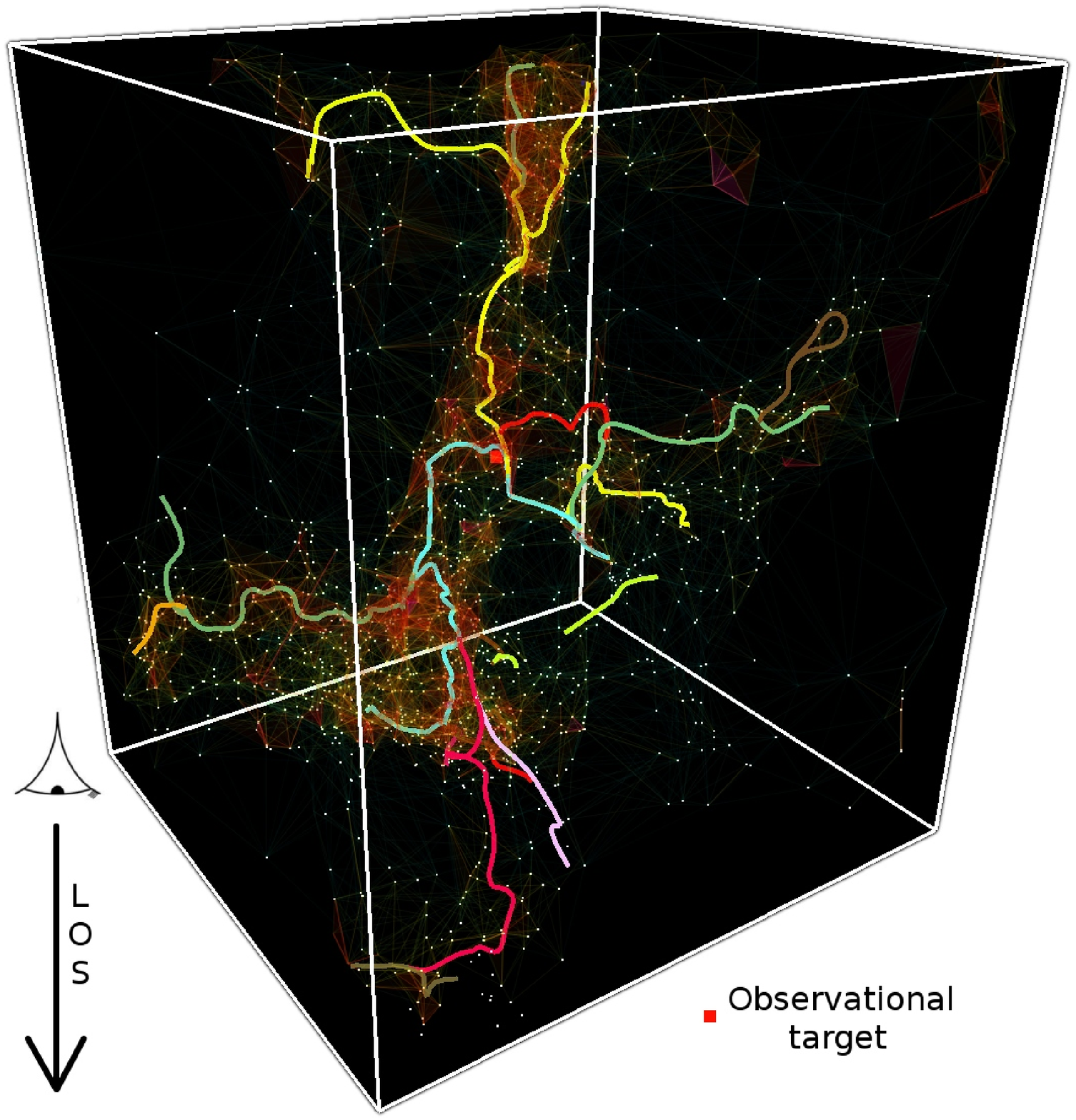}
\end{minipage}
\caption{Left: An X-ray halo observed around an elliptical galaxy in the center of a group at redshift $z=0.083$ and located at the confluence of several filaments. The color map indicates the X-ray combined image of CCD chips (XIS 0,1, and 3), while the white dots stand for the SDSS spectroscopic-identified galaxies within $0.080<z<0.086$. The filamentary structure in the surrounding region is shown by the colored solid curves, extracted from the filaments catalogue shown on figure \ref{fig_SDSS_picB}. Note that the colors (cyan, red and yellow) correspond to that of the filaments represented on the 3D view on the right frame. Right: a 3D view of the configuration of the filaments around the observed region. The vertical axis corresponds to the line of sight (the observer being upward), and the box roughly encompasses the galaxies in the SDSS catalogue with coordinates $233^\circ<{\rm DEC}<243^\circ$, $22^\circ<{\rm RA}<32^\circ$ and $0.075<z<0.092$. The delaunay tesselation of the galaxies, shaded according to the local density, is displayed to help visualizing the filamentary structure. The observational target is identified by a red square and is located at the intersection of the red  cyan and yellow filaments, the last two  being aligned with the line of sight to a very good approximation. A movie is available for download at {\em\tt http://www.iap.fr/users/sousbie/}. \label{fig_observation} }
\end{figure*}

\subsection{An ``optically faint'' cluster at a filamentary junction}
Because some dark matter haloes are sparsely populated and also as a result of selection effects, classical methods such as FOF are unable to detect them from the observed galaxy distribution. Such ``optically faint'' groups and clusters may nevertheless present a strong astrophysical interest: as they are by nature different from the ``regular'' haloes, one could for instance expect that they have different formation history that needs to be understood. As they are faint though, their properties are poorly assessed, but massive dark matter haloes such as galaxy clusters or galaxy groups are believed to form at intersections of two or several filaments, which can be identified in the SDSS using \progname. We demonstrate that this is possible by enlightening the relationship between an X-ray halo and its surrounding filamentary network as identified in the SDSS catalogue (see figure \ref{fig_SDSS_picB}).\\

 Because of the particular configuration of the filaments in the region, we submitted an observation proposal to the X-ray satellite SUZAKU \citep{2007PASJ...59S...1M}, which was accepted. We present here the results of this observation, but reserve its analysis to a future article (Kawahara {\it et al} 2010, {\it in prep.}). The observational target was selected for being located at the confluent of galaxy filaments, and because one of those filaments is both straight and aligned with the line of sight as shown on figure \ref{fig_observation} (see the yellow filament on the right frame). While no X-ray signal could be found within the ROSAT All Sky Survey (RASS), X-ray signals emitted by diffuse thermal gas were clearly observed by the high sensitivity detectors of SUZAKU, unveiling the presence of a dark matter halo as shown by the X-ray image reproduced in the central part of the left panel of figure \ref{fig_observation}. It is remarkable that there are no corresponding candidates in the $78,800$ groups catalogue identified by \citep{2010A&A...514A.102T} using a modified friend-of-friend (FOF) algorithm.  In fact, because the optically observable member galaxies are not strongly clustered and their number is limited ($N \sim 10$), regular methods have high chances to miss them. It is also very difficult to locate and identify particular filamentary configurations by eye directly from the galaxy distribution using projections or even a real time 3D visualization. Using \progname, we showed that it is possible to easily identify such targets, which demonstrate the complimentary of our approach with respect to one based on a traditional halo finders.      

\section{Significance of topology of LSS}
\label{subsec_sigthres}
As noted in paper I, it is not an option to use the raw  \hyperref[defDMC]{Discrete Morse-Smale complex}  as a tool to assess the properties of the cosmic web. Hence we showed there  how to simulate a topological simplification of the DTFE density field so that the critical simplexes that were most probably accidentally generated by Poisson noise could practically be removed from the \hyperref[defDMC]{DMC}. This simplification is based on the \hyperref[defpers]{persistence} ratio of critical points pairs (\ie \hyperref[defpers]{persistence} pairs), and one must therefore decide a significance level $s=\nsig{n}$ such that all \hyperref[defpers]{persistence} pairs with lower significance (\ie or equivalently a higher probability to be generated by Poisson noise) can be removed. We showed in paper I that, at least in the 2D case, such a method allows for what seems to be a very efficient and natural simplification of the \hyperref[defDMC]{DMC}. We did not discuss however how to decide the value of this particular threshold. This is particularly important though, and especially in the context of the cosmic web, as our ultimate goal is to assess physical properties of astrophysical objects identified as features of the \hyperref[defDMC]{DMC} (\ie the haloes, filaments, walls and voids of matter distribution on cosmological scales in the Universe). Imagine for instance one is interested in statistically measuring the average number of filaments that branch on dark matter halos. If the threshold is too low, the measure will be equivalent to that in a Gaussian random field because of Poisson noise (see lower left frame of figure 13 of paper I), and if it is too high, then the risk is to systematically ignore weaker filaments (see central right panel of figure 13 of paper I).\\

\begin{figure}
\centering\includegraphics[width=\linewidth]{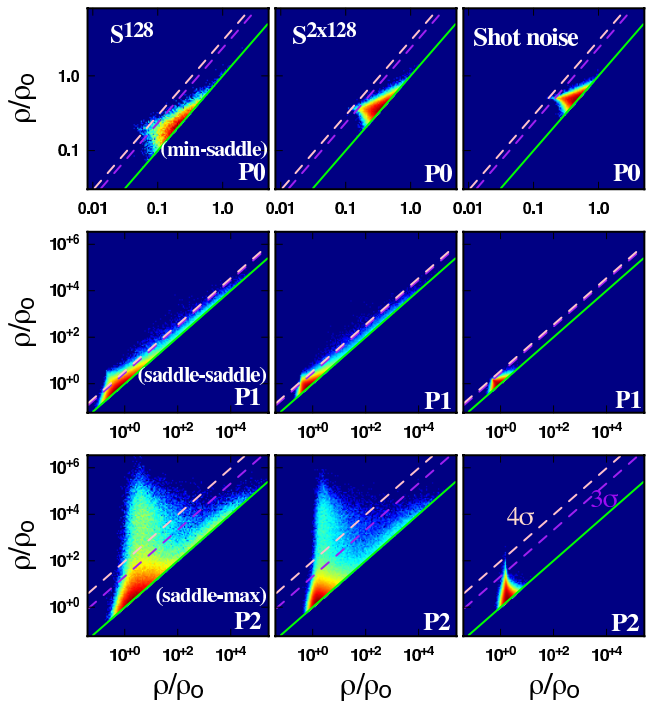}
\caption{\hyperref[defpers]{persistence} diagrams (\ie the probability distribution function (PDF) of \hyperref[defpers]{persistence} pairs) in a cosmological simulation and for Gaussian random noise. Each pair $P_i=\ppair{p_i}{q_{i+1}}$ of critical points of order $i$ and $i+1$ is considered as a point with coordinates $[\rho\left(p_i\right),\rho\left(q_{i+1}\right)]/\rho_0$. The PDF were computed from a $250\Mpc$ large $\Lambda$CDM dark matter simulation down sampled to $128^3$ particles, $S^{128}$ (left column), the same distribution with $128^3$ additional randomly located particles, $S_N^{2\times 128}$ (central column), and a random distribution of particles within the same volume, $S_R^{128}$ (right column). From top to bottom, each line correspond to a different type of pair: $P_0$ (minima/$2$-saddle points), $P_1$ ($2$-saddle points/$1$-saddle points) and $P_2$ ($1$-saddle points/maxima) respectively. The green, purple dashed and pink dashed lines correspond to $\nsig{0}$, $\nsig{3}$ and $\nsig{4}$ \hyperref[defpers]{persistence} levels respectively.\label{fig_per_diag}}
\end{figure}

\begin{figure}
\centering\includegraphics[width=\linewidth]{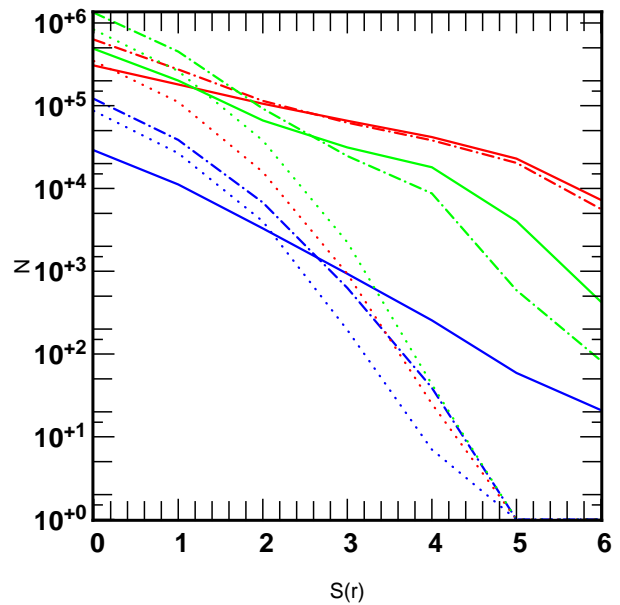}
\caption{Number of \hyperref[defpers]{persistence} pairs of type $k$ as a function of the significance threshold $S_k\left(r\right)$ (in units of $\sigma$) in a $250\Mpc$ large $\Lambda$CDM dark matter simulation down sampled to $128^3$ particles, $S^{128}$ (filled curves), the same distribution with $128^3$ additional randomly located particles, $S_N^{2\times 128}$ (dash-dotted curve) and a random distribution of particles within the same volume, $S_R^{128}$(dotted curves). The blue, green and red color correspond to \hyperref[defpers]{persistence} pairs of type $0$, $1$ and $2$ respectively (see figure \ref{fig_per_diag} for the corresponding \hyperref[defpers]{persistence} diagrams).\label{fig_pair_count_simu}}
\end{figure}
\subsection{Persistence diagrams}
Figure \ref{fig_per_diag} shows the probability distribution function of \hyperref[defpers]{persistence} diagrams (see \citet{edel00}, \citet{cohen07}) computed from the Delaunay tesselation of a $250\Mpc$ large, $512^3$ particles $\Lambda$CDM dark matter simulation subsampled to $128^3$ particles (left column, $S^{128}$ hereafter), the same distribution with an identical number of particle added at random locations (central column, $S_N^{2\times 128}$ hereafter), and a completely random distribution of particles within the same volume (right column, $S_R^{128}$ hereafter). Simply speaking, plotting a \hyperref[defpers]{persistence} diagram of a density distribution $\rho$ basically consists in representing each \hyperref[defpers]{persistence} pair $P_i=\ppair{p_i}{q_{i+1}}$, where $p_i$ and $q_{i+1}$ are critical points of order $i$ and $i+1$ respectively, as a point of coordinates $[\rhod,\rhou]=[\rho\left(p_i\right),\rho\left(q_{i+1}\right)]/\rho_0$ where $\rho_0$ designates the average density in the distribution\footnote{In the following, the term density will generally refer to the normalized density $\rho/\rho_0$ so that different distributions can be fairly compared}. Recall that \hyperref[defpers]{persistence} pairs are pairs of critical simplexes that correspond to the act of creation and destruction of a topological feature in the \hyperref[deffiltration]{Filtration} of the Delaunay tesselation. 
 On figure \ref{fig_per_diag}, the pairs of type $P_0$, $P_1$ and $P_2$ are represented on the top, central and bottom rows respectively. On those diagrams, the pairs with null \hyperref[defpers]{persistence} lie on the green line of equation $\rhou=\rhod$ and the farther away from this line a point is, the higher the \hyperref[defpers]{persistence} of its corresponding \hyperref[defpers]{persistence} pair. The purple and pink dashed line stand for $\nsig{3}$ and $\nsig{4}$ \hyperref[defpers]{persistence} respectively. As expected, most \hyperref[defpers]{persistence} pairs in the random distribution $S_R^{128}$ have a \hyperref[defpers]{persistence} ratio below $\nsig{3}$ (right column). Fortunately, the PDF of the \hyperref[defpers]{persistence} pairs in $S^{128}$ is sufficiently different from that in $S_R^{128}$ so that a reasonable fraction of them lie above the $\nsig{3}$ and even $\nsig{4}$ threshold (see left column). By canceling all those pairs that lie below the $\nsig{3}$ or $\nsig{4}$ line, it should therefore seem reasonable to assume that only those topological properties that were imprinted by the physical processes at work in the simulation would be conserved. A good measure of the actual influence of Poisson noise on the distribution of the \hyperref[defpers]{persistence} pairs in the underlying distribution can be gained from the examination of the central column. The distribution $S_N^{2\times 128}$ was created by adding a large number of randomly located particles to $S^{128}$, resulting also in the creation of a very large number of spurious critical points. One can see on the central column that as a result, the \hyperref[defpers]{persistence} diagram tends to concentrate at lower \hyperref[defpers]{persistence} ratio (\ie closer from the green line). This  means that as expected, those spurious critical points mainly create low \hyperref[defpers]{persistence} ratio pairs which can therefore be removed.\\

This observation is supported by figure \ref{fig_pair_count_simu}, where the actual number of \hyperref[defpers]{persistence} pairs in the three distributions are displayed as a function of the cutting threshold. Whereas the number of critical pairs of all sorts and with significance higher than $\nsig{0}$ is higher in $S_N^{2\times 128}$ (dash-dotted curves) than in $S^{128}$ (plain curves), this number decreases comparatively faster with the increase of the \hyperref[defpers]{persistence} selection threshold. For low \hyperref[defpers]{persistence} thresholds (\ie up to $\sim \nsig{2}$), the number of \hyperref[defpers]{persistence} pairs in $S_N^{2\times 128}$ actually decreases as fast as that in the random distribution $S_R^{128}$ (dotted curves). In the case of pairs of type $P_1$ and $P_2$ ($2$-saddle points/$1$-saddle points pairs, green curves, and $1$-saddle point/maxima pairs, red curves, respectively), this tendency actually changes between $\nsig{2}$ and $\nsig{3}$ and the cancellation rates in $S_N^{2\times 128}$ and $S^{128}$ become relatively similar above $\nsig{3}$. This strongly suggests that most of the spurious \hyperref[defpers]{persistence} pairs in $S_N^{2\times 128}$ do in fact have a \hyperref[defpers]{persistence} ratio lower than $\nsig{3}$ and that above that threshold, the remaining \hyperref[defpers]{persistence} pairs have a distribution similar to that in the original N-body simulation $S^{128}$. The \hyperref[defpers]{persistence} pairs of type 0 in $S_N^{2\times 128}$ ( minima/$2$-saddle point pairs, blue filled curves) exhibit a slightly different behavior though, as their number seems to vary more or less accordingly with the \hyperref[defpers]{persistence} threshold in $S_N^{2\times 128}$ and $S_R^{128}$ (blue dotted curve). This number nevertheless always remain higher in $S_N^{2\times 128}$ and there are proportionally more high \hyperref[defpers]{persistence} pairs in $S_N^{2\times 128}$ than in $S_R^{128}$. This  suggests that the number of minima resulting from the physical processes at stake in voids formation is relatively low compared to that due to Poisson noise, the reason for this being that the cosmological voids' minima have an intrinsically lower density because of the nature of voids. While Poisson noise creates spurious minima over a wide range of densities, the voids' minima only span the lower densities and therefore stretch over comparatively larger scales due to DTFE properties (resolution being inversely proportional to the density). The addition of random particles in $S_N^{2\times 128}$ particularly affects the wider regions around minima, therefore increasing their density and lowering the \hyperref[defpers]{persistence} ratio of the corresponding \hyperref[defpers]{persistence} pairs, hence the lack of high significance pairs of type $0$ at $S\left(r\right)>\nsig{5}$ (see blue curves) in $S_N^{2\times 128}$ compared to $S^{128}$. Note however that this does not mean that the physically created \hyperref[defpers]{persistence} pairs are destroyed by Poisson noise in $S_N^{2\times 128}$, but only that they are shifted to lower \hyperref[defpers]{persistence}, and that the \hyperref[defpers]{persistence} threshold should not be chosen too high if ones wants to retrieve the full \hyperref[defDMC]{DMC} (which is not the case if one is only interested in the filaments).\\

\begin{figure*}
\begin{centering}
\subfigure[Critical points PDF]{\includegraphics[width=0.49\linewidth]{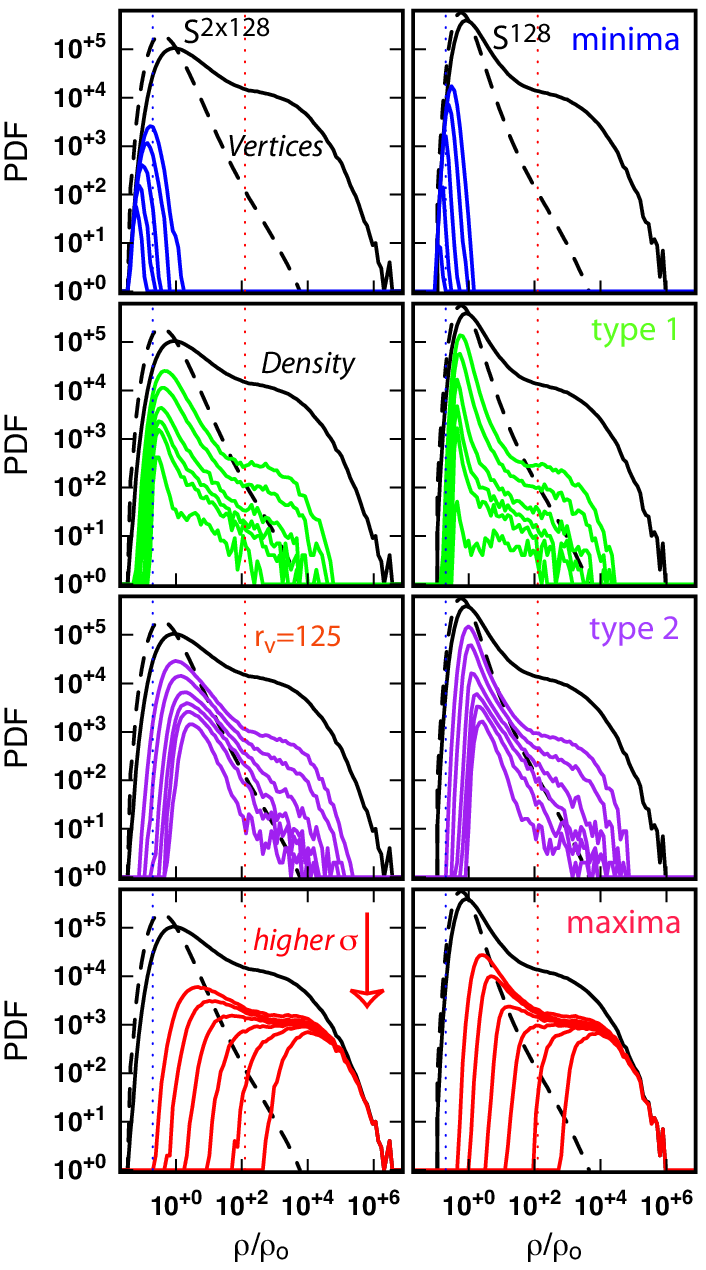}\label{fig_simu_crit}}\hfill
\subfigure[Betti numbers and Euler characteristic]{\includegraphics[width=0.49\linewidth]{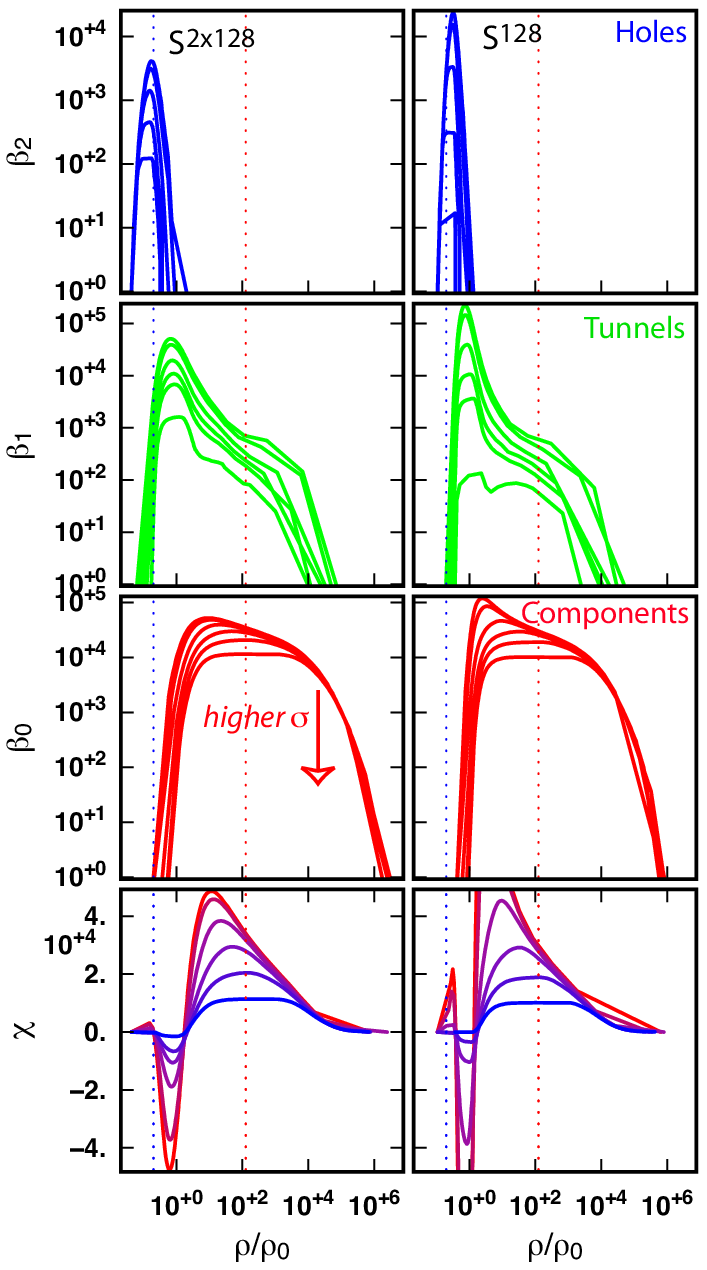}\label{fig_simu_betti}}
\end{centering}
\caption{Evolution of the topological properties in a $512^3$ particles $250 \Mpc$ dark matter simulation down-sampled to $128^3$ particles, $S^{128}$, for increasing \hyperref[defpers]{persistence} levels (left columns on each figure), and in the same distribution with $128^3$ additional randomly located particles, $S_N^{2\times 128}$ (right columns on each figure). On each frame, the \hyperref[defpers]{persistence} selection level ranges from $\nsig{0}$ for the outer colored curve up to $\nsig{6}$ for the inner curve. {\em Left:} The probability distribution function (PDF) of critical points of type $0$ (top) up to $3$ (bottom) as a function of their overdensity $\rho/\rho_0$. The black curve is the PDF of the vertice in the tessellation while the dashed curve stands for the (volume weighted) PDF of the overdensity $\rho/\rho_0$. The blue and red vertical dotted lines emphasize the critical level $r_v=\rho_v/\rho_0=0.2$ (resp. $r_p=\rho_p/\rho_0=125$) below (resp. above) which a void (resp. a peak) may be considered physically significant. {\em Right:} from top to bottom, the betti numbers, $\beta_2$, $\beta_1$, $\beta_0$, and Euler characteristic $\chi$ of the excursion set with over density greater than $\rho/\rho_0$. \label{fig_simu_curves}}
\end{figure*}

Two complementary measures of the evolution of the topological properties in $S^{128}$ and $S_N^{2\times 128}$ with the \hyperref[defpers]{persistence} threshold are presented on figure \ref{fig_simu_curves}: the PDF of the critical points on figure \ref{fig_simu_crit} and the betti numbers and Euler characteristics on figure \ref{fig_simu_betti}.

\subsection{Critical points}

 Let us consider figure \ref{fig_simu_crit} first. On that figure, the PDF of the density at vertice (\ie the particles in the studied distribution) is shown by the dark black bold curve, and it is striking how the PDF of the critical points tend to follow it, especially at low \hyperref[defpers]{persistence} (outer curves): the more the $k$-simplexes at a given density level, the higher the number of detected critical points of order $k$. This is an expected result when Poisson noise dominates as it affects indifferently any scales, but it is not desirable though as the filamentary structure of the cosmic web is an intrinsic property which should not depend on the properties of a particular sampling technique. One would in fact rather expect the PDF of the critical points to follow the PDF of the volume weighted density, or equivalently as we use DTFE, of the number of vertice at a given density in the tesselation\footnote{in the case of DTFE, the density of a sample particle is defined as the inverse volume of its dual Voronoi cell, and the volume it occupies is also the volume of this cell, which implies that the PDF of the volume weighted density and that of the number of sample particles are identical.}. The black bold dashed curve traces the {\em volume weighted} PDF of the density at vertice. It is clear on figure \ref{fig_simu_crit} that in the case of the minima, $1$-saddle points and $2$-saddle points  PDFs, the bias toward higher better sampled densities due to DTFE is progressively wiped out with increasing \hyperref[defpers]{persistence} ratio threshold, and almost disappears above a significance level threshold of $\sim\nsig{3}$ (see blue, green and purple curves). The PDF of the maxima though (red curves) exhibits an opposite tendency, as their PDF  concentrates at higher and higher densities with increasing \hyperref[defpers]{persistence} ratio thresholds. This actually reflects the nature of the distribution of the dark matter over large scales in the universe. In fact, most maxima are expected to be found within gravitationally bound structures undergoing non-linear regime (\ie dark matter haloes), which therefore exhibit densities several order of magnitudes higher than the average density and with very steep gradients (note that this fact also prevents them from being affected by Poisson noise too much). Those regions, although numerous, represent only a very small fraction of the total volume, as reflected by the discrepancy between the PDF of the maxima at high \hyperref[defpers]{persistence} ratio and the volume weighted PDF of the density. To confirm these hypothesis, we traced on figures \ref{fig_simu_curves} and \ref{fig_simu_betti} the blue and red vertical dotted lines which mark the characteristic average under-density of a void in a Einstein-de Sitter model, $\rho/\rho_0\leq 0.2$ (see \citet{void_blumenthal92}, \citet{void_sheth04} or \citet{neyrinck}) and the typical critical overdensity above which gravitationally bound structure are identified using friend of friend algorithm, $\rho/\rho_0\geq 125$ \citep{FOFinfo} respectively. While this is not clear at low \hyperref[defpers]{persistence} thresholds because of Poisson noise, all maxima (resp. minima) belonging to \hyperref[defpers]{persistence} pairs with \hyperref[defpers]{persistence} ratio greater than $\sim\nsig{3}$ have densities above (resp. below) those critical thresholds while the two types of saddle points lie within those limits. This  means that the detected persistent maxima and minima correspond to physically meaningful objects, which strongly supports the pertinence of using \hyperref[defpers]{persistence} based cancellation of a Morse-Smale complex to identify the characteristics components of the cosmic web such as cosmic voids and filaments.\\

\subsection{Discrete topological invariants}

The Betti numbers and Euler characteristics represented on figure \ref{fig_simu_betti} are slightly more involved topological analysis tools than the PDF of critical points (see paper I for a more formal definition of the Betti numbers and a simple example of their computation). The $k^{\rm th}$ Betti number $\beta_k$  counts the number of \hyperref[defkcycle]{$k$-cycles} in excursion sets as a function of the density threshold of the excursion. Within the context of the 3D cosmological matter distribution, there are $3$ Betti numbers, that count the number of holes or $2$-cycles ($\beta_2$), the number of tunnels or $1$-cycles ($\beta_1$) and the number of distinct components or $0$-cycles ($\beta_0$) enclosed in the set of points with density threshold larger than the aforementioned density threshold. As this threshold decreases, new \hyperref[defkcycle]{$k$-cycles} may be created or destroyed, therefore increasing or decreasing the value of the corresponding Betti numbers. The value of the Betti numbers as a function of the density threshold reflects the global topology of the field (\ie the way it connects as function of density threshold) and it is therefore very instructive to compare the Betti numbers of two distributions to appreciate how similar or distinct they may be from a topological point of view. For that reason, we plotted on figure \ref{fig_simu_betti}, from top to bottom, the value of $\beta_2$, $\beta_1$, $\beta_0$ and the Euler characteristic $\chi$ (a topological invariant, computed as the alternate sum of the Betti numbers) as measured in $S^{128}$ and $S_N^{2\times 128}$ (left and right column respectively). As noted in \paperone, the notions of \hyperref[defpers]{persistence} pairs and Betti numbers are intimately related: the Betti numbers were readily computed from the \hyperref[defpers]{persistence} pairs, the positive critical point of order $k+1$ increasing $\beta_k$ when it enters the excursion and the negative critical point of order $k$ decreasing $\beta_k$. Distribution  $S_N^{2\times 128}$ was obtained by adding an equal number of randomly distributed particles to the particles in the N-body simulations $S^{128}$, and the Betti numbers of the two distributions should therefore give some insight on how topology is affected by Poisson noise. Note that the presence of Poisson noise in $S_N^{2\times 128}$ affects the PDF of the sampled density by slightly downscaling it (numerous random particles land in large scale void regions, increasing their densities, while few of them affect the high density regions, therefore lower their density contrast, see black plain curves on figure \ref{fig_simu_curves}). When comparing Betti numbers in the two distributions, one would rather want to know weather the same structures (\ie void, tunnel, component) exist in both distributions though, even if it exists at slightly different densities. It is therefore more important to compare the general shape and amplitude of the Betti number in both distributions than their value at a precise density threshold. Inspecting figure \ref{fig_simu_betti}, it is clear that random particles mainly affect the topological properties of the field around the average density $\rho_0$, each Betti number differing of about one order of magnitude in $S^{128}$ (left) and $S_N^{2\times 128}$ (right) at a level around $\rho/\rho_0=1$. The situation largely improves after the cancellation of the lower \hyperref[defpers]{persistence} pairs though and it is striking how the shape and amplitude of the Betti numbers at a level of \hyperref[defpers]{persistence} ratio of $3\sim\nsig{4}$ become similar. Note also that $\beta_0$ is the Betti number that is the least affected by Poisson noise, and for \hyperref[defpers]{persistence} higher than $\nsig{3}$, the values are almost identical in $S^{128}$ and $S_N^{2\times 128}$. This  means that individual components in the \hyperref[deffiltration]{Filtration} are created and merge in a very similar way independently of the presence of Poisson noise, which does not affect the filamentary structure of $S^{128}$. It is therefore reasonnable to trust the filaments detected at \hyperref[defpers]{persistence} levels higher than $\sim\nsig{3}$ as being true topological properties of the underlying distribution. One should nonetheless remain cautious with the identification of voids and wall. In fact, although the topology of the $1$-cycles and $2$-cycles seems to be correctly recovered in $S_N^{2\times 128}$ at a significance level of $3\sim\nsig{4}$, this is not the case anymore at higher levels and one should be careful not to set the threshold too high. In fact, the cosmological voids and walls are more affected by Poisson noise as they usually live at densities around $\rho/\rho_0=1$ where the influence of Poisson noise is maximal and the corresponding \hyperref[defpers]{persistence} pairs have statistically lower \hyperref[defpers]{persistence} ratios than that associated to filaments.\\

\section{Conclusion}
\label{sec_conclusion}

We implemented \progname\,  (Soubie  2010)  on realistic 3D dark matter cosmological simulations and
observed  redshift catalogues from the SDSS DR7.
 We showed that \progname\, traces  very well the observed filaments, walls, and voids seen both in simulations and observations.
In either setting, filaments are shown to connect onto halos, outskirt  walls, which circumvent voids, as is topologically required  by Morse theory.  
Indeed, \progname\,  warrants that all the well-known and extensively studied mathematical properties of Morse theory are ensured by construction at the mesh level.
As illustrated in sections~\ref{sec_SDSS}, \progname\, assumes nothing about the geometry of the survey  or its homogeneity, and yields a natural (topologically motivated)
 self-consistent criterion for selecting the significance level of the identified structures.
We demonstrated   that the extraction  is possible even for very sparsely sampled  point processes,  as a function of the \hyperref[defpers]{persistence} ratio 
(a measure of the significance of topological  connections between  critical points), which allows us to account consistently for the shot noise of real surveys.
The corresponding recovered cosmic web is also ``persistent" in as much as it is robust because it relies on intrinsic topological features of the underlying density field. 
% .
Hence we can now trace precisely the locus of filaments, walls and voids from such samples and  assess the  confidence of the post-processed sets  as a function of this threshold,
which can be expressed relative to the expected amplitude of  shot noise.
\progname\, also seems to be robust, in as much that more sparsely samples recover filamentary structures which are consistent with those of the better sampled catalogues.
In a cosmic framework, this criterion was shown to level  with FoF structure finder for the identifications of peaks, while \progname\,  also identifies the connected filaments and quantitatively produces on the fly the  full set of Beti numbers (number of holes, tunnels, connected components etc...) {\sl directly from the particles}, as a function of the \hyperref[defpers]{persistence} threshold, as these follow from the \hyperref[defpers]{persistence} pairs. 
We investigated the evolution of the critical points, the  Beti numbers and the Euler characteristic has a function of the \hyperref[defpers]{persistence} ratio: its illustrates the biases
 involved in filtering low \hyperref[defpers]{persistence} ratios.
For dark matter simulations, this criterion was shown to be sufficient even if one particle out of two is noise,  when the  \hyperref[defpers]{persistence} ratio is set to 3-$\sigma$ or more. 
We applied this procedure to the localization of a specific filamentary configuration and observed an ``optically faint''  cluster at a galaxy filaments junction, identified in the SDSS catalogue. An X-ray counterpart could indeed be observed (Kawahara et al. in prep) by the X-ray satellite SUZAKU.  The filaments of the SDSS extracted with \progname\,  are available online at the URL {\em \tt http://www.iap.fr/users/sousbie/SDSS-skeleton.html} as a set of segments with extremities in RA, DEC, redshift. All these results  are very encouraging for future investigations using  \progname, for searching galaxy clusters, galaxy groups, and missing baryons of the universe in particular, and for the study of LSS in general.

\subsection*{Acknowledgements}
{

\sl The authors thank T.~Nishimichi for his help dealing with SDSS galaxy catalogue, H.~Yoshitake for his help with X-ray analysis and TS thanks Y.~Suto for his constant help and support. 

This work was made possible through an extensive usage of the Yorick programming language by D.~Munro (available at {\em\tt http://yorick.sourceforge.net/}) and also CGAL, the Computational Geometry Algorithms Library, ({\em\tt http://www.cgal.org}),  to compute the Delaunay tessellations.

\sl The filaments of the SDSS extracted with \progname\, is readily available online at the URL {\em \tt http://www.iap.fr/users/sousbie/SDSS-skeleton.html}.

\sl Funding for the SDSS has been provided by the Alfred P. Sloan Foundation, the Participating Institutions, the National Science Foundation, the U.S. Department of Energy, the National Aeronautics and Space Administration, the Japanese Monbukagakusho, the Max Planck Society, and the Higher Education Funding Council for England. The SDSS Web Site is {\em\tt http://www.sdss.org/}. 
\sl TS gratefully acknowledges support from JSPS (Japan Society for the Promotion of Science) Postdoctoral Fellowhip for Foreign Researchers award P08324. 
 HK is supported by a JSPS (Japan Society for Promotion of Science) Grant-in-Aid for science fellows.
 CP acknowledges supports from a  Leverhulme visiting professorship at the Astrophysics department of the University of Oxford.
  This work is also supported by Grant-in-Aid for Scientific research from JSPS and from the Japanese Ministry of Education, Culture, Sports, Science and Technology (No. 22$\cdot$5467).

}

\bibliographystyle{mn2e}
\bibliography{morse-illustation}
%\cleardoublepage

%\clearpage
%\pagebreak
%\newpage
%\onecolumn

\appendix
\cleardoublepage
\onecolumn
%\newpage

\section*{Terminology}
\label{sec_terminology}
%\descriptionlabel{$\bullet$}
\description
\item[{\bf Arc}]\label{defarc} An arc is a $1$-cell: an integral line (or a V-path in the discrete theory) whose origin and destinations are critical points. The arcs of \hyperref[defMSC]{Morse-Smale complex}  connect two critical points of order difference $1$ (\ie in 2D, a minimum and a saddle-point or a maximum and a saddle-point).   
\item[{\bf $n$-cell}]\label{defcell} A $n$-cell is a region of space of dimension $n$ such that all the integral lines in the $n$-cell have a common origin and destination. The $n$-cells basically partition space into regions of uniform gradient flow 
\item[{\bf Coface}] \label{defcoface} A coface of a $k$-simplex $\alpha_k$ is any $p$-simplex $\beta_p$, with $p\geq q$, such that $\alpha_k$ is a face of $\beta_p$. In 3D, the cofaces of a segment (\ie a $1$-simplex) are any triangle or tetrahedron (\ie $2$ or $3$-simplex) whose set of summits (\ie vertexes) contains the two vertexes at the extremities of the segment, as well as the segment itself.
\item[{\bf Cofacet}]\label{defcofacet}  A cofacet of a $k$-simplex $\alpha_k$ is a coface $\beta_{k+1}$ of $\alpha_k$ with dimension $k+1$. Equivalently, $\alpha_k$ is a facet of $\beta_{k+1}$.
\item[{\bf Critical point of order $k$}] For a smooth function $f$, a critical point of order $k$ is a point such that the gradient of $f$ is null and the Hessian (matrix of second derivatives) has exactly $k$ negative eigenvalues. in $2$D, a minimum, saddle point and maximum are critical points of order $0$, $1$ and $3$ respectively.
\item[{\bf Critical $k$-simplex}] A critical $k$-simplex is the equivalent in discrete Morse theory of the critical point of order $k$ in its smooth counterpart. Note that in $2$D, the equivalent of a minimum is a critical vertex ($0$-simplex), a saddle-point is a critical segment ($1$-simplex) and a maximum is a critical triangle ($2$-simplex). 
\item[{\bf Crystal}]\label{defcristal} A crystal is a $3$-cell: a 3D region delimited by $6$ quads and $12$ arcs, within which all the integral lines (or V-pathes in the discrete case) have identical origin and destinations.
\item[{\bf $k$-cycle}]\label{defkcycle}  A $k$-cycle in a simplicial complex corresponds to a $k$ dimensionnal topological feature. in $3D$, $0$-cycles correspond to independant components, $1$-cycles to loops and $2$-cycles to shells
\item[{\bf Discrete Gradient}]\label{defDG} A discrete gradient of a discrete Morse-Smale function $f$ defined over a simplicial complex $K$ pairs simplexes of $K$. Within a gradient pair, the simplex with lower value is called the tail and the other the head, and any unpaired simplex is critical.
\item[{\bf Discrete Morse-Smale complex (DMC)}]\label{defDMC} The discrete Morse-Smale complex (DMC for short) is the equivalent of the Morse-Smale complex applied to simplicial complexes.
\item[{\bf Discrete Morse-Smale function}]\label{defDMF} A discrete Morse-Smale function $f$ defined over a simplicial complex $K$ associates a real value $f\left(\sigma_k\right)$ to each simplex $\sigma_k\in K$ .
\item[{\bf Excursion set}] see sub-level set.
\item[{\bf Face}]\label{defface} A face of a $k$-simplex $\alpha_k$ is any $p$-simplex $\beta_p$ with $p\leq q$, such that all vertexes of $\beta_p$ are also vertexes of $\alpha_k$. In 3D, the faces of a $3$-simplex (\ie a tetrahedron) are the tetrahedron itself, the $4$ triangles that form its boundaries, the $6$ segments that form its edges, and its $4$ summits (\ie vertexes). 
\item[{\bf Facet}]\label{deffacet} A facet of a $k$-simplex $\alpha_k$ is a face $\beta_{k-1}$ of $\alpha_k$ with dimension $k-1$. The facets of a $3$-simplex (\ie a tetrahedron) are the $4$ triangles (\ie $2$-simplexes) that form its boundaries.
\item[{\bf Filtration}]\label{deffiltration} A filtration of a simplicial complex $K$ is a {\em growing} sequence of sub-complexes $K_i$ of $K$, such that each $K_i$ is also a simplicial complex. If the different $K_i$ are defined by a discrete function $F_\rho$ as the set of simplexes of $K$ with value $F_\rho\left(\sigma\right)$ less or equal to a given threshold, a filtration can be though of as the discrete equivalent of a sequence of growing sub-level sets of a smooth function.
\item[{\bf Gradient pair / arrow}]\label{defGP} A Gradient pair or arrow is a set of two simplex, one being the facet of the other, and such that they are paired within a discrete gradient. Within a gradient pair, the simplex with lower value is called the tail and the other the head.
\item[{\bf Integral line}] An integral line of a scalar function $\rho\left({\mathbf x}\right)$ is a curve whose tangent vector agrees with the gradient of $\rho\left({\mathbf x}\right)$.  
\item[{\bf Level set / Sub-level set)}] A level set, also called iso-contour, of a function $\rho\left({\mathbf x}\right)$ at level $\rho_0$ is the set of points such that $\rho\left({\mathbf x}\right)=\rho_0$. The corresponding Sub-level set is the set of points such that $\rho\left({\mathbf x}\right)\geq\rho_0$ 
\item[{\bf Ascending/Descending $p$-manifold}]\label{defmanifold} Within a space of dimension $d$, an ascending $p$-manifold is the set of points from which, following minus the gradient, one reaches a given critical point of order $d-p$. A descending $p$-manifold is the set of points from which, following the gradient, one reaches a given critical point of order $p$. For istance, ascending $1$-manifolds in 3D can be associated to the filaments, and ascending $3$-manifolds describe the voids 
\item[{\bf Morse function}]\label{defMF} A Morse function is a continuous, twice differentiable smooth function whose critical points are non degenerate. In particular the eigenvalues of the Hessian matrix (\ie the matrix of the second derivatives) must be non-null 
\item[{\bf Morse complex}] The Morse complex of a Morse function is the set of its its ascending (or descending) manifolds.
\item[{\bf Morse-Smale function}] A Morse-Smale function is a Morse function whose ascending and descending manifolds intersect {\em transversely}. This means that there exist no point where an ascending and a descending manifold may be tangent 
\item[{\bf Morse-Smale complex}]\label{defMSC} The Morse-Smale complex is the intersection of the ascending and descending manifolds of a Morse-Smale function. One can think of the Morse-Smale complex as a network of critical points connected by $n$-cells, defining a notion of hierarchy and neighborhood among them. In particular, the geometry of the arcs (\ie $1$-cells) is determined by the critical integral lines (\ie integral lines that join critical points) and the order of two critical points connected by an arc may only differ by $1$.  
\item[{\bf Peak/Void patch}] In 3D, a peak patch is a descending $3$-manifold (\ie the region of space from which, following the gradient, one reaches a given maximum), and a void patch an ascending $3$-manifold (\ie the region of space from which, following minus the gradient, one reaches a given minimum). 
\item[{\bf Persistence}] \label{defpers} The persistence of a persistence pair (or equivalently of the corresponding $k$-cycle it creates and destroys) is defined as the difference between the value of the two critical points (or critical simplexes in the discrete case) in the pair. It basically represents its life time within the evolving sub-level sets (or filtration in the discrete case). 
\item[{\bf Persistence pair}]\label{defperspair} In the smooth context of a function $\rho$, persistence pairs critical points $P_a$ and $P_b$ of $\rho$ that respectively create and destroy a topological feature (or $k$-cycle) in the sub-level sets of $\rho$, at levels $\rho\left(P_a\right)$ and $\rho\left(P_b\right)$. In the discrete case of a simplicial complex $K$, a persistence pair is a pair of critical simplexes $\sigma_a$ and $\sigma_b$ of a given discrete function $F_\rho\left(\sigma\right)$, such that $\sigma_a$ creates a $k$-cycle (\ie topological feature) when it enters the filtration of $K$ according to $F_\rho$ and $\sigma_b$ destroys it when it enters. 
\item[{\bf Persistence ratio}] \label{defpersR} The persistence ratio of a persistence pair (or equivalently of the corresponding $k$-cycle it creates and destroys) is the ratio of the value of the two critical points (or critical simplexes in the discrete case) in the pair. Persistence ratio is preferred to regular persistence in the case of strictly positive functions such as the density field of matter on large scales in the universe.  
\item[{\bf Quad}] \label{defquad}A quad is a $2$-cell : a 2D region delimited by four arcs within which all the integral lines (or V-pathes in the discrete case) have identical origin and destinations.
\item[{\bf $k$-simplex}] \label{defksimplex}A $k$-simplex is  the $k$ dimensional analog of a triangle: the simplest geometrical object with $k+1$ summits, called vertex. It is the building block of simplicial complexes 
\item[{\bf Simplicial complex}] A simplicial complex $K$ is a set of simplexes such that if a $k$-simplex $\alpha_k$ belongs to $K$, then all its faces also belong to $K$. Moreover, the intersection of two simplexes in $K$ must be a simplex that also belongs to $K$ 
\item[{\bf Vertex}] A vertex is a $0$-simplex or simply a point.
\item[{\bf V-path}] \label{defVP}A V-path is the discrete equivalent of an integral line: it is a set of simplexes linked by discrete gradient arrows and face/coface relation. Tracing a V-path  consists in intuitively following the direction of the gradient pairs of a discrete gradient from a critical simplex to another. 
\enddescription
%\end{description}

\label{lastpage}

% define DMC, discrete morse complex
% define simplicial complex
% define Morse function
 % define filtration

\end{document}